# Moments of the Counts Distribution
# in the 1.2 Jy IRAS Galaxy Redshift Survey [1]

FRANÇOIS R. BOUCHET

*Institut d'Astrophysique de Paris, CNRS,*

*98 bis Boulevard Arago, F-75014 Paris, FRANCE*

MICHAEL A. STRAUSS

*Institute for Advanced Study, School of Natural Sciences*

*Princeton, New Jersey 08540*

MARC DAVIS

*Astronomy and Physics Departments, University of California*

*Berkeley, California 94720*

KARL B. FISHER

*Institute of Astronomy, Madingley Rd., Cambridge CB3 0HA, England*

AMOS YAHIL

*Astronomy Program, State University of New York*

*ESS Building, Stony Brook, New York 11794–2100*

JOHN P. HUCHRA

*Harvard-Smithsonian Center for Astrophysics, 60 Garden Street*

*Cambridge, Massachusetts 02138*

1

**Abstract**

We have measured the count probability distribution function (CPDF) in a series of 10 volume-limited sub-samples of a deep redshift survey of *IRAS* galaxies. The CPDF deviates significantly from both the Poisson and Gaussian limits in all but the largest volumes. We derive the volume-averaged 2, 3, 4, and 5-point correlation functions from the moments of the CPDF, and find them all to be reasonably well-described by power laws. Weak systematic effects with the sample size provide evidence for stronger clustering of galaxies of higher luminosity on small scales. Nevertheless, remarkably tight relationships hold between the correlation functions of different order. In particular, the "normalized" skewness defined by the ratio $S_3 \equiv \overline{\xi_3}/\overline{\xi_2}^2$ varies at most weakly with scale in the range $0.1 < \overline{\xi_2} < 10$. That is, $S_3$ is close to constant ($= 1.5 \pm 0.5$) from weakly to strongly non-linear scales. On small scales, this is consistent with previous determinations of the three-point correlation function $\zeta \equiv \xi_3$. On larger scales, this conforms with the hypothesis of the growth of observed structures by gravitational clustering of initially Gaussian density fluctuations. We similarly find that $\overline{\xi_4}$ is proportional to the third power of $\overline{\xi_2}$ in the same range of $\overline{\xi_2}$, and there is weak evidence that $\overline{\xi_5}$ is proportional to the fourth power of $\overline{\xi_2}$. Furthermore, we find that the void probability function obeys a scaling relation with density to great precision, in accord with the scale-invariance hypothesis ($\overline{\xi_N} \propto \overline{\xi_2}^{N-1}$). Double-counting cluster galaxies in order to match the cluster overdensities seen in optically selected samples of galaxies increases greatly the derived value of $S_3$ and $S_4$, although the scaling between the the correlations of different orders remains. Unfortunately, the relative sparseness of the *IRAS* sample preclude using it to make the most demanding tests of scale invariance, which rely on the overall shape of the CPDF at different scales. In this sparse limit, various models for the CPDF become degenerate, and fit the *IRAS* data nearly equally well. Indeed, the CPDF is well fitted by both the negative binomial distribution, and the thermodynamical model of Saslaw and Hamilton, and to a somewhat lesser extent by the log-normal distribution. All three models fit the data poorly for the densest subsample of *IRAS* galaxies examined, but this may be more a reflection of finite volume effects than of the inadequacy of the models.

## 1. Introduction

The two-point correlation function of the galaxy distribution, and its Fourier transform, the power spectrum, have long been the principle statistical tools by which astronomers have quantified the clustering of galaxies (Peebles 1980 and references therein). However, as redshift surveys have revealed ever more complex structures in the distribution of galaxies, the need for statistics which illuminate other aspects of the galaxy distribution has become acute. For example, the large voids recently discovered in the galaxy distribution (de Lapparent, Geller, & Huchra 1986; Kirshner *et al.* 1987; Geller & Huchra 1989) are not mirrored by any feature in the two-point correlation function, and their description requires a rather different statistic. On small scales ($\lesssim 500$ km s$^{-1}$), one observes clusters of galaxies which represent enormous overdensities, and whose properties are only completely described by the full complement of $N$-point correlation functions up to values of $N$ equal to the number of galaxies in the cluster. In practice, standard techniques have great difficulty measuring correlation functions of order four and higher from finite samples (Peebles 1980). Another approach is that taken by Szapudi, Szalay, & Boshan (1992) and Meiksin, Szapudi, & Szalay (1992) (see also Szapudi & Szalay 1993), who relate counts in cells to higher-order correlations, enabling them to go to $N = 8$ in the angular correlation function.

In this paper, we examine various properties of the counts of galaxies in cells of a given size $\ell$. In particular, we define the quantity $P_N(\ell)$ as the fraction of randomly positioned spheres of radius $\ell$ containing exactly $N$ galaxies, for a given galaxy sample. As is clear from the definition, $P_N$ depends strongly on the mean density of galaxies in a sample, a property which it does not share with the $N$- point correlation function. However, there may exist scaling relations, which we present below, which allow us to express the $P_N$ in density-independent ways.

The early work on counts in cells is summarized by Peebles (1980), where its connection to the correlation function is made (cf., White 1979). In particular, the void probability function can be related to an infinite sum of the $N$-point correlation functions $\xi_N$:

$$P_0(\ell) = \exp\left[\sum_{N=1}^{\infty} \frac{(-\overline{N})^N}{N!} \overline{\xi_N}(v)\right] \quad , \tag{1}$$

where $\overline{N}$ is the expected number of galaxies in the absence of clustering in the volume $v \equiv (4\pi/3)\ell^3$, and $\overline{\xi_N}(v)$ is the volume average of the correlation function:

$$\overline{\xi_N}(v) \equiv \frac{1}{v^N} \int_v d^3\mathbf{r}_1 d^3\mathbf{r}_2 \ldots d^3\mathbf{r}_N \, \xi_N(\mathbf{r}_1, \mathbf{r}_2, \ldots, \mathbf{r}_N) \quad . \tag{2}$$

Note that in an unclustered universe, ($\xi_N = 0$ for $N > 1$), the void probability function is given by the Poisson expression

$$P_0(\ell) = \exp(-\overline{N}) \quad . \tag{3}$$

The probability that a cell of volume $v$ contains $N$ galaxies, $P_N(\ell)$, is directly related to $P_0$: the void probability function clearly depends on the average density of galaxies $n \equiv \overline{N}/v$, and one can derive from Eq. 1:

$$P_N(\ell) = \frac{(-n)^N}{N!} \frac{\partial^N P_0(n,\ell)}{\partial n^N} \quad . \tag{4}$$

For example, inserting Eq. 3 in Eq. 4 yields

$$P_N(\ell) = \frac{\overline{N}^N}{N!} e^{-\overline{N}} \quad , \tag{5}$$

which is just the Poisson distribution, as it should be in the unclustered limit.

Based on early results for the three- and four-point correlation functions in the non-linear regime, a number of workers hypothesized the existence of a scaling hierarchy, in which the high-order correlations could be expressed as symmetrized sums of lower-order correlation functions (cf., Fry & Peebles 1978; Fry 1984; Schaeffer 1984; Sharp, Bonometto, & Lucchin 1984). The subject reached maturity with the papers of Balian & Schaeffer (1988; 1989a; 1989b) who start with a generic assumption of scale invariance for the correlation function hierarchy:

$$\xi_N(\lambda \mathbf{r}_1, \cdots \lambda \mathbf{r}_N) = \lambda^{-\gamma(N-1)} \xi_N(\mathbf{r}_1, \cdots \mathbf{r}_N) \quad , \tag{6}$$

for any value of $N$ over some (large) range of scales. This happens, for instance, if the $\xi_N$ are proportional to a symmetric product of $N-1$ two-point correlation functions, as is suggested by the observed form of the three- and four-point correlation functions in the non-linear regime ($\xi_2 \gtrsim 1$). In any case, it implies that the volume-averaged correlation functions satisfy:

$$\overline{\xi_N}(v) = S_N \, \overline{\xi_2}^{N-1}(v) \quad , \tag{7}$$

where the $S_N$ are independent of scale (although different scale invariant systems or different scale ranges may have different sets of values of the $S_N$). An immediate consequence of this follows by inserting Eq. 7 into Eq. 1: if we write

$$P_0(\ell) = \exp[-\overline{N}\sigma(n,v)] \quad , \tag{8}$$

where $\sigma$ quantifies the deviation of $P_0$ from the Poisson prediction (Eq. 3 above), then $\sigma$ is a unique function of the quantity $N_c$, given by

$$N_c \equiv \overline{N}\,\overline{\xi_2}(v) \quad. \tag{9}$$

If $\xi_2$ is a power-law of index $\gamma$, one has (Peebles & Groth 1976)

$$\overline{\xi_2} = J_2\,\xi_2 \equiv \frac{72\xi_2}{(3-\gamma)(4-\gamma)(6-\gamma)\,2^\gamma} \quad, \tag{10}$$

while the average number of neighbors of a galaxy in a sphere of radius $r$, minus the number in a homogeneous universe, is

$$N_{cluster} = n\int_0^r \xi_2\,d^3\mathbf{r} = 3\overline{N}\overline{\xi_2}/(3-\gamma) \quad, \tag{11}$$

or

$$N_c = \frac{24}{(4-\gamma)(6-\gamma)\,2^\gamma}N_{cluster} \approx 0.77 N_{cluster} \text{ for } \gamma = 1.8 \quad. \tag{12}$$

Thus $N_c(\ell)$ is an indicator of the typical number of clustered particles on the scale $\ell$.

If scale invariance (Eq. 6) holds, Eqs. (1) and (4) imply a number of scaling relations of the count probability distributions (Balian & Schaeffer 1989a) which involve two universal functions governing the behavior of the $P_N$. A particularly clear and succinct summary of these relations is given in Bouchet, Schaeffer, & Davis (1991; hereafter BSD), who find that they are well-satisfied in simulations of a universe dominated by Cold Dark Matter. Bouchet & Hernquist (1992) find similar results for a white noise initial spectrum, although numerical limitations do not allow them to make as firm a statement for Hot Dark Matter. In any case, these studies seem to indicate that scale invariance is a generic result of the gravitational instability process applied to gravitating matter, when density contrasts become large. Observationally, the scale-invariant predictions have been found to hold for the galaxy distribution in the Center for Astrophysics redshift survey (CfA) (Alimi, Blanchard, & Schaeffer 1990) and the Southern Sky Redshift Survey (SSRS) (Maurogordato, Schaeffer, & da Costa 1992) (see also Gaztañaga 1992). Unfortunately, the limited volume of these surveys has not allowed all aspects of the scale-invariant predictions to be tested over a large range in $\overline{\xi_2}$. In particular, as we will see below, the most stringent tests of the scale-invariant hypothesis occur at the highest sampling densities, which existing redshift surveys probe in only very small volumes.

There is not as yet any dynamical theory explaining the onset of scale invariance in the fully non-linear regime, although self-similar solutions of the BBGKY hierarchy do exist (Davis & Peebles 1977). However, when density contrasts are weak, one can use a perturbative approach to study the early stages of the process. The statistical properties of a Gaussian random density field, (i.e., one in which the phases of its different Fourier modes are uncorrelated), are completely described by the power spectrum, which is the Fourier conjugate of the two-point correlation function; all higher-order (reduced) correlations are zero. In linear theory, the growth of density perturbations under gravitational instability preserves the Gaussian nature of the density field. However, it was recognized long ago that even when density contrasts are small, non-linearities induce deviations from Gaussianity. Bernardeau (1992) has shown that in the weakly non-linear regime (i.e., as long as $\overline{\xi_2} \ll 1$), gravity applied to an initially Gaussian density field induces the correlation hierarchy $\overline{\xi_N} = S_N\overline{\xi_2}^{N-1}$ for all $N \geq 1$, where the $S_N$ are independent both of scale and of the initial power spectrum of density fluctuations. This generalizes to arbitrary $N$ the result already obtained for $S_3$ by Peebles (1980), for $S_4$ by Fry (1984), and for $S_5$ by Goroff et al. (1986). Bouchet et al. (1992b) have furthermore shown analytically that, in this weakly non-linear regime, $S_3$ is insensitive to the value of the density parameter $\Omega_0$, and is only weakly affected by the mapping from real space to redshift space, although both $\overline{\xi_2}$ and $\overline{\xi_3}$ change. Juszkiewicz,

Bouchet, & Colombi (1993a) show how the convolution of the density field with a Gaussian or spherical top-hat window (appropriate for the counts-in-cells analyses we use here) introduces a dependence of $S_3$ on the local slope of the power spectrum. They further show how the comparison of the variance and skewness of the galaxy distribution can be used to distinguish between models with Gaussian distributed initial density perturbations, and models with exotic seeds such as textures, for which one would generically expect $\overline{\xi_3} \propto \overline{\xi_2}^{3/2}$. An introduction to these latter topics may be found in Juszkiewicz & Bouchet (1992). Of course, there is no *a priori* reason to believe that the correlation hierarchy established on a given scale during the mildly non-linear stages of galaxy evolution will survive the later strongly non-linear stages; in addition, one might expect mode coupling between small and large scales (cf., Little, Weinberg, & Park 1991). However, results of numerical simulations by Bouchet & Hernquist (1992), Weinberg & Cole (1992), Lahav *et al.* (1993), and Fry, Melott, & Shandarin (1993) have indeed shown that measurements at large scale obey the perturbative theory, even when the systems have reached full non-linearity at small scales. The self-similar BBGKY solutions imply that scale invariance can be generated on very non-linear scales, but there is no *a priori* connection between the values of $S_N$ determined on large and small scales, and there is no reason to assume that they are the same.

In this paper, we examine the nature of the $P_N$ as derived from volume-limited subsamples of a redshift survey of galaxies detected by the *Infrared Astronomical Satellite* (*IRAS*). The sample consists of 5304 galaxies with 60 micron flux density above 1.2 Jy, selected over 87.6% of the sky. The selection criteria for the galaxies are given in Strauss *et al.* (1990) and Fisher (1992), and the data for the brighter half of the sample are given in Strauss *et al.* (1992b). *IRAS* galaxies are a dilute tracer of the galaxian density field (Strauss *et al.* 1992a), with typically 1/3 the number density of galaxies appearing in optically selected samples of comparable depth. Thus this paper emphasizes those properties of the counts distribution that are accessible in the low-density limit; the large volume covered by our sample allows many independent volumes of a given size at a given number density to be probed. In particular, we explore the relationships between the counts in cells and the correlation functions, and are able only to start to probe the asymptotic forms that the count distributions take in the limit of large density (cf. BSD, and Colombi, Bouchet & Schaeffer 1993b). A preliminary version of this work was presented in Bouchet, Davis, & Strauss (1992).

The outline of the paper is as follows. In § 2, we discuss the measurement of $P_N$ from the *IRAS* data. In § 3 the various moments of the counts distribution are derived from the $P_N$. The correlation functions $\overline{\xi_N}$ are presented in § 3.1. In § 3.2, we explore the relationship between correlation functions of different order. The scaling of the void probability function with density is shown in § 4, and the $P_N$ are compared with various scale-invariant models. We discuss the results and summarize in § 5.

## 2. The Sample and its Analysis

We select from the redshift sample ten volume-limited subsamples, each containing roughly twice the volume of the previous one. The outer radius, number of galaxies included, and the corresponding minimum luminosity of each subsample are given in Table 1. Observed heliocentric redshifts are corrected to the barycenter of the Local Group using the correction of Yahil, Tammann & Sandage (1977). No correction is made to the redshifts for peculiar velocities. Our rationale for this is two-fold: Bouchet *et al.* (1992b) tell us that the value of $S_3$ is insensitive to peculiar velocities in second-order perturbation theory, and our method for self-consistently correcting galaxy redshifts for peculiar velocities (Yahil *et al.* 1991) is unable to properly model small-scale features in the velocity field (Davis, Strauss, & Yahil 1991), which could cause unknown effects on the counts in cells analysis. Thus proper distances $r$ are estimated via their measured redshift $z$, according to

$$H_0 r = \frac{2cz(z + 2 - \Omega_0)}{(1+z)(1+(1+\Omega_0 z)^{1/2})(1-\Omega_0+(1+\Omega_0 z)^{1/2})} \quad . \tag{13}$$

The proper distances corresponding to each outer radius are listed in Table 1. Luminosities are assigned to galaxies as proportional to $f_{60} r^2$ in order to determine their membership in any given volume-limited sample; thus we ignore the $K$-correction in calculating luminosity (cf., the discussion in Fisher *et al.* 1992; Fisher *et al.* 1993b). The results presented here were obtained with $\Omega_0 = 1$. Fisher *et al.* (1992) found no evidence for number density evolution in this sample; we thus ignore this possible effect. We adopt a value of $H_0 = 100 \,\mathrm{km\,s^{-1}\,Mpc^{-1}}$ throughout this paper.

Three further series of volume-limited samples were created to test the sensitivity of the counts to various effects. In the first, we used Eq. 13 with $\Omega_0 = 0.1$; this makes negligible differences in the results. In the second, we used $\Omega_0 = 1$, but placed all galaxies in the seven clusters in Table 2 of Yahil *et al.* (1991) at a common redshift to collapse the fingers of God associated with each. We show below the effect this has on the moments. Finally, Strauss *et al.* (1992a) shows that the galaxy densities in cores of clusters determined from *IRAS* galaxies are systematically lower than those determined from optically selected galaxies; with this in mind, we give galaxies associated with cluster cores a weight given by the ratio of the seventh and fifth columns of Table 2 of Strauss *et al.* (1992a). We refer to the counts derived from this analysis as the boosted counts.

Within each volume-limited subsample, we place down $10^6$ points at random, and count the number of galaxies within a series of concentric spheres around this point. We only count spheres which are completely included within the subsample volume, and which do not intersect the largest region of sky uncovered by the survey, namely that at $|b| < 5°$. In addition, four percent of the high-latitude sky is uncovered by the survey (Strauss *et al.* 1990); we fill these regions with random points at the same number density as the observed galaxies. The number of random galaxies in each subsample is indicated in the last column of Table 1. The large number of randomly placed spheres makes the *measurement* error of the counts negligible. The *statistical* errors due to the finite volume of the samples is non-negligible, however, and will be discussed further in the next section.

Fig. 1 shows the resulting counts in cells for the subsamples volume-limited to 2400 km s$^{-1}$ (top), 6000 km s$^{-1}$ (middle), and 12,000 km s$^{-1}$ (bottom). If the galaxies are unclustered, the $P_N$ are given by the Poisson distribution, Eq. 5. We do not show this case in the figure to reduce the clutter; suffice it to say that the curves deviate strongly from this limit. A more interesting case is to assume that the second moment of the distribution describes it in full. In the limit that the higher-order moments are negligible, we might approximate the $P_N$ as a Gaussian:

$$P_N = \frac{1}{(2\pi \overline{N}^2 \mu_2)^{1/2}} \exp\left[-\frac{(N-\overline{N})^2}{2\overline{N}^2 \mu_2}\right] \quad , \tag{14}$$

where $\mu_2$ is the second moment of the counts, given in Eq. 15 below. This Gaussian expression is shown as dots in the figure. Note that Eq. 14 expression is *not* that one would get by assuming that the $\overline{\xi_N}$ are negligible for $N > 2$ in Eq. 1. In any case, the figure shows that Eq. 14 is a poor fit to the data. Notice however, that for the larger scales (the curves which peak at larger $N$ in the figure), the relative importance of the higher-order correlations drops, and the Gaussian expression becomes a better approximation. Moreover, in the sparsely sampled limit (lowest panel of the figure), the higher-order correlations become less important, and again the Gaussian approximation becomes better.

Our volume-limiting strategy means that a fraction of the galaxy data is unused (compare the numbers in Table 1 with the 5304 galaxies of the full survey), which is of course not optimal. However, most galaxies in the sample fall into at least one of the subsamples. Two possible alternatives are the following: one could divide the sample into a series of shells around the observer, each considered as a sub-sample of a given average number density. One would then proceed to measure the counts in cells inside each shell, ignoring the effect of the number density gradient on the scale of the shell (Efstathiou *et al.* 1990). Alternatively, one could weight each galaxy of the full sample by the inverse of the selection function to obtain a constant average number density irrespective of distance to the observer, and proceed as in a volume-limited sample (Saunders *et al.* 1991). But this raises the issue of assessing the effect of a varying noise level on the scale of the cell for the statistics we might consider. In the present paper, we have taken a cautious approach; by treating galaxies of different luminosities separately, we avoid introducing any bias other than those that might be due to the sample selection alone.

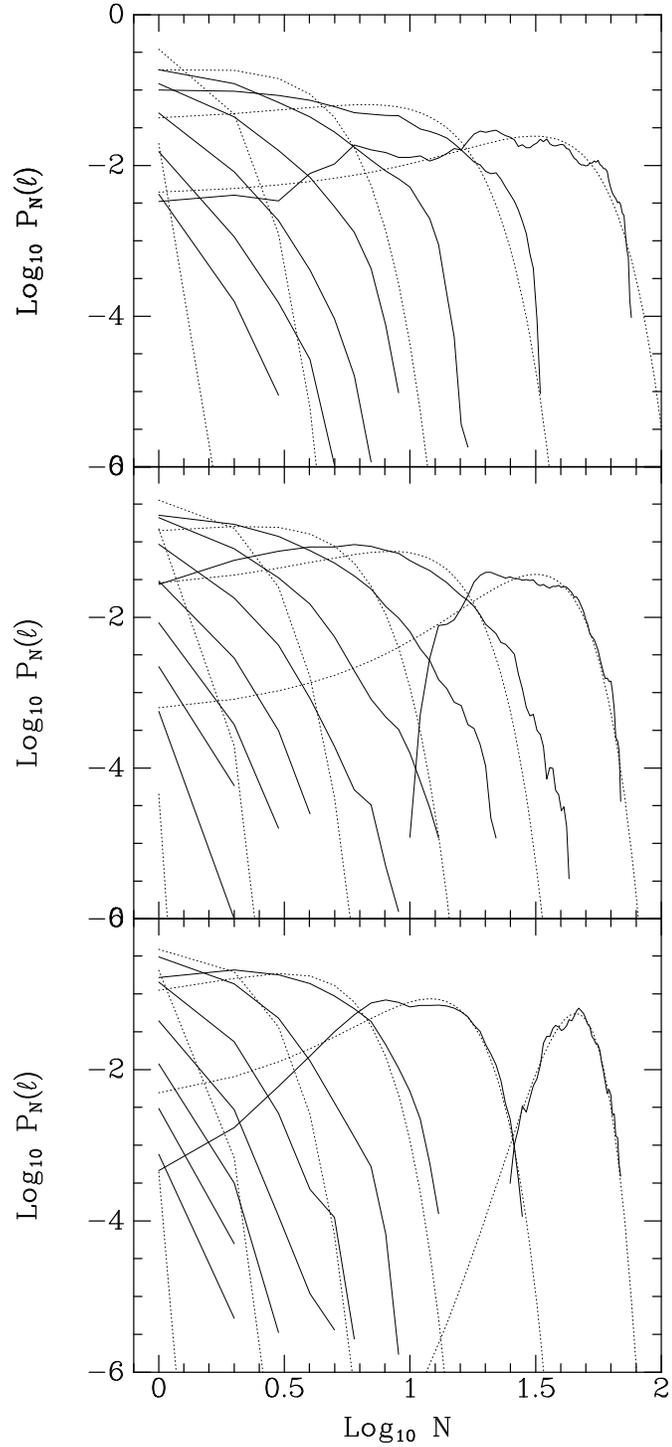

Figure 1: Measured Count probabilities $P_N(\ell)$ (solid), in the 2400 km s$^{-1}$ (top), 6000 km s$^{-1}$ (middle) and 12,000 km s$^{-1}$ (bottom) volume-limited sub-samples, as a function of the count number $N$ for different cell sizes (radii) $\ell$: from bottom to top the scales $\ell$ are 0.5, 0.8, 1.3, 2.0, 3.2, 5.0, 7.9, 12.6, and 19.9 $h^{-1}$ Mpc. The dots show a gaussian distribution with the same variance as the data.

## 3. The Moments of the Count Distribution

There are two approaches to measuring the correlation functions of galaxies from the counts in cells. The first works directly from the expressions relating the two types of statistics (cf., Eq. 1), and is explained in an Appendix. We find it to be of limited applicability, however, and here we relate the correlation function to the moments of the counts distribution. Once we know the $P_N(\ell)$, we can compute various centered moments of the distribution

$$\mu_M(\ell) = \left\langle \left(\frac{N - \overline{N}}{\overline{N}}\right)^M \right\rangle = \sum_{N=0}^{\infty} \left(\frac{N - \overline{N}}{\overline{N}}\right)^M P_N(\ell) \quad, \tag{15}$$

where $\overline{N} \equiv \langle N \rangle = \sum N P_N(\ell)$. Thus $\overline{N}$ is the average density of galaxies in a subsample, times $v$. The volume-averaged correlation functions are the irreducible moments (*e.g.*, Peebles 1980), which equal zero for a Gaussian distribution. They are given by

$$\overline{\xi_2}(\ell) = \mu_2 - \frac{1}{\overline{N}}, \quad \overline{\xi_3}(\ell) = \mu_3 - 3\frac{\mu_2}{\overline{N}} + \frac{2}{\overline{N}^2} \quad, \tag{16}$$

$$\overline{\xi_4}(\ell) = \mu_4 - 6\frac{\mu_3}{\overline{N}} - 3\mu_2^2 + 11\frac{\mu_2}{\overline{N}^2} - \frac{6}{\overline{N}^3} \quad, \tag{17}$$

and

$$\overline{\xi_5}(\ell) = \mu_5 - 10\frac{\mu_4}{\overline{N}} - (10\mu_2 - \frac{35}{\overline{N}^2})\mu_3 + 30\frac{\mu_2^2}{\overline{N}} - 50\frac{\mu_2}{\overline{N}^3} + \frac{24}{\overline{N}^4} \quad, \tag{18}$$

where we defined the average correlation functions of order $N$ over the cell volume $v = (4\pi/3)\,\ell^3$ in Eq. 2. For $N = 2, 3$, and 4, these approach the variance, skewness, and kurtosis of the distribution in the continuum limit, $\overline{N} \to \infty$. In the following we refer to these terms in that limit.

Note that the contribution of the densest fluctuations increases with the order, making high orders increasingly sensitive to rare high-density peaks. Thus, for example, $\overline{\xi_5}$ will depend strongly on the nature of the richest clusters, which are rare enough that their number in the finite volume probed will be dominated by Poisson fluctuations. One can in principle correct for this if the asymptotic behavior of $P_N$ at large $N$ is known or can be fitted for (cf., Colombi & Bouchet 1992; Colombi, Bouchet, & Schaeffer 1993a). However, the $P_N$ reach their asymptotic forms only in the densely sampled regime (Colombi *et al.* 1993b), which even the smallest volume-limited subsample of *IRAS* galaxies does not approach. Thus it is quite difficult in practice to apply finite volume corrections to the moments derived from the counts in cells in the *IRAS* survey, and we do not attempt to do so here.

As definition (2) shows, the $\overline{\xi_N}$ are volume averages of the standard correlation functions. Although the volume averaging procedure discards all detailed geometrical information in $\xi_N$, it enhances the signal to noise ratio, and preserves a fundamental quantity, namely the clustering strength as a function of scale. Thus it allows us to explore the scaling properties of the statistics; given the data sets we have at present, this might be the best we can do.

As we shall see, the sparseness of the *IRAS* sample prevents us from probing the deeply non-linear regime in detail. But we can study the critical transition region between the linear and non-linear regimes, over at least 2 decades in scale, and more than 3 decades in the value of the variance $\overline{\xi_2}$, so that both domains are partly accessible.

*3.1. Volume-Averaged Correlation Function*

Fig. 2 shows the determinations of $\overline{\xi_2}(\ell)$ resulting from Eq. 16, for all the volume-limited subsamples of Table 1. There are systematic effects with the size of the sample. The samples drawn from the smallest volume do not allow any measurement of the correlations at large scales. As the sample volume increases, more independent structures are included in the volume, and the correlations on large scales increase, until they stabilize when the ratio of sample volume to $v \equiv (4\pi/3)\ell^3$ becomes large enough, just as expected from finite volume effects (Colombi *et al.* 1993a). At the smallest scales, $\overline{\xi_2}$ tends to increase with the sample volume (although no determination is possible for the largest volumes). This is probably evidence for a weak luminosity effect, in the sense that the more luminous *IRAS* galaxies in the larger volume-limited subsamples are slightly more clustered than less luminous galaxies. Davis *et al.* (1988) and Yahil *et al.* (1991) concluded that there were no *gross* effects as a function of luminosity, but their tests were somewhat less sensitive than those here. Fisher *et al.* (1993b) also concluded that the correlation function was independent of luminosity, but the smallest volume for which they calculated the volume-limited correlation function was to 6000 km s$^{-1}$, by which point the effects have largely disappeared. For each subsample, we make a least-squares fit of the points to the form $\log_{10} \overline{\xi_2} = A_2 + B_2 \log_{10} \ell$; the results are given in Table 2 and Fig. 6 below. The effects we are seeing here are small, however; as Davis *et al.* argue, the stability of the correlation function derived from different volumes argues against the hypothesis that the galaxy distribution is described by a naïve pure fractal (e.g., Pietronero 1987; cf., the discussion in Peebles 1993).

We average the determinations of $\overline{\xi_2}(\ell)$ of Fig. 2 at each separation with equal weights, yielding the squares in Fig. 3; the error bars are the standard deviation from the mean. The volume averaged correlation function $\overline{\xi_2}(\ell)$ is quite well described by a single power-law over more than 2 decades in scale from $\overline{\xi_2}(\ell) \sim 40$ at $\ell \simeq 0.5\, h^{-1}$ Mpc down to a value as small as $\overline{\xi_2}(\ell) \sim 0.03$ at $\ell \simeq 50\, h^{-1}$ Mpc, with no significant sign of a "break" or a "bump" at large scale. A least squares fit to the data yields (short dashes) $\overline{\xi_2}(\ell) = A_2 + B_2 \log_{10} \ell$, with $A_2 = 1.17 \pm 0.05$ and $B_2 = -1.59 \pm 0.06$ (corresponding to $\ell_0 = 5.44 \pm 0.53\, h^{-1}$ Mpc where $\overline{\xi_2}(\ell_0) \equiv 1$). Because the points in Fig. 3 are correlated with one another, it is not strictly correct to make a least-square fit to $\overline{\xi_2}$ (cf., the discussion in Fisher *et al.* 1993b). However, the deviations from our least squares fit are reasonably Gaussian, as may be judged from the distribution of deviations from our power-law fit plotted in Fig. 4. The quoted error bars are thus representative of the dispersion of the data.

Fisher *et al.* (1993b) derive the correlation function $\xi(s)$ of the full *IRAS* sample using the standard methods of pair counts. Because $\xi(s)$ is not a pure power law, we cannot use Eq. 10 to relate their results to ours; rather, we fit a spline to $\xi(s)$ and numerically integrate Eq. 2 to find $\overline{\xi_2}$; the results are shown as solid triangles in the figure. The two approaches to the correlation function agree perfectly on large scales, but on smaller scales ($r < 3$ Mpc), the counts in cells analysis shows a systematically larger correlation function. In particular, the counts in cells analysis gives a near-perfect power law, while the pair counts correlation function become more and more shallow at smaller separations. The Fisher *et al.* correlation function is derived from the full *IRAS* sample with optimal weighting, which means that the correlation function at small scales is dominated by counts from nearby and therefore low-luminosity galaxies, while our equal weight averaging gives more weight to more luminous galaxies. The lower luminosity galaxies show weaker clustering, as we saw above, which causes the turnover of the correlation function on small scales. Indeed, the triangles in Fig. 3 are in perfect agreement with $\overline{\xi_2}$ determined from the volume-limited sample to 2400 km s$^{-1}$, shown as open circles.

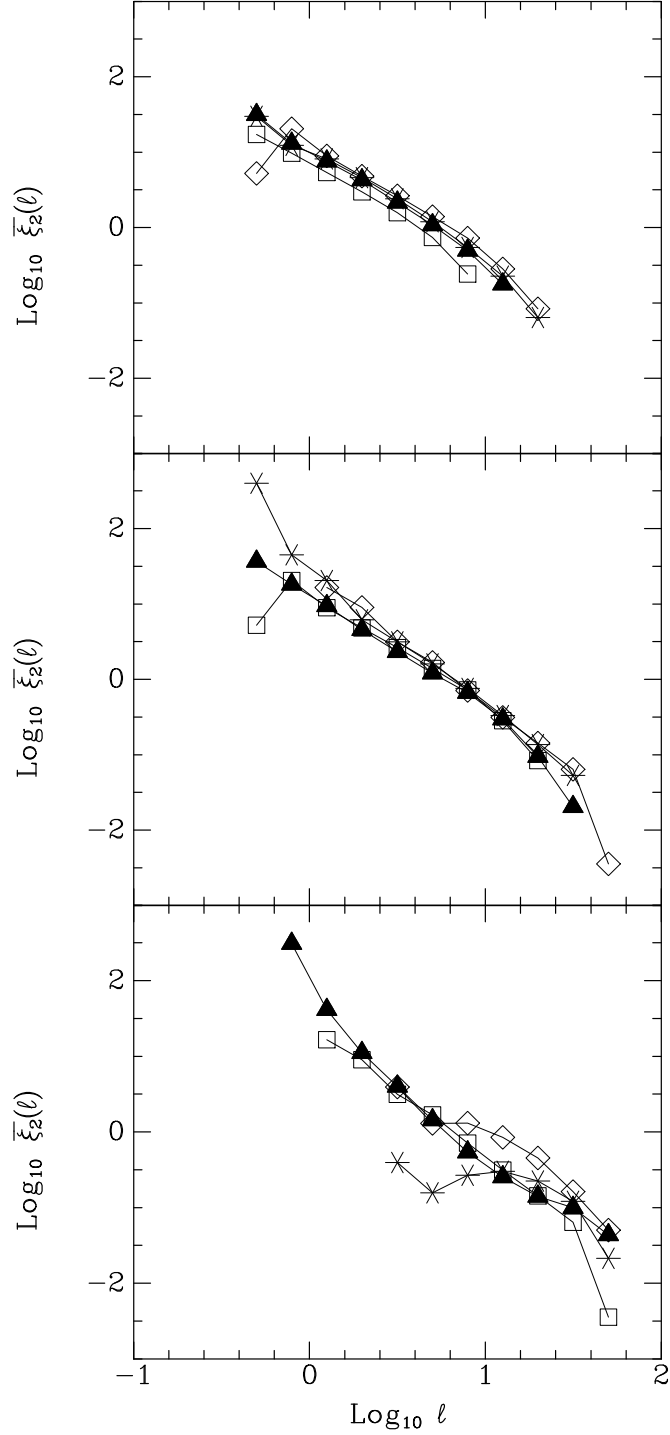

Figure 2: Raw determinations of the two-body correlation function $\overline{\xi_2}(\ell)$, for all 10 volume-limited subsamples. The top panel shows the subsamples limited to 2400, 3900, 4800, and 6000 km s$^{-1}$. The middle one shows the 6000, 7600, 9500, and 12,000 km s$^{-1}$ subsamples, while the bottom panel displays the largest subsamples corresponding to 12,000, 15,000, 19,000, and 24,000 km s$^{-1}$. In each panel, the open squares are the correlation function determined from the smallest volume, the triangles from the next volume, the stars from the next volume, and the open diamonds from the largest volume.

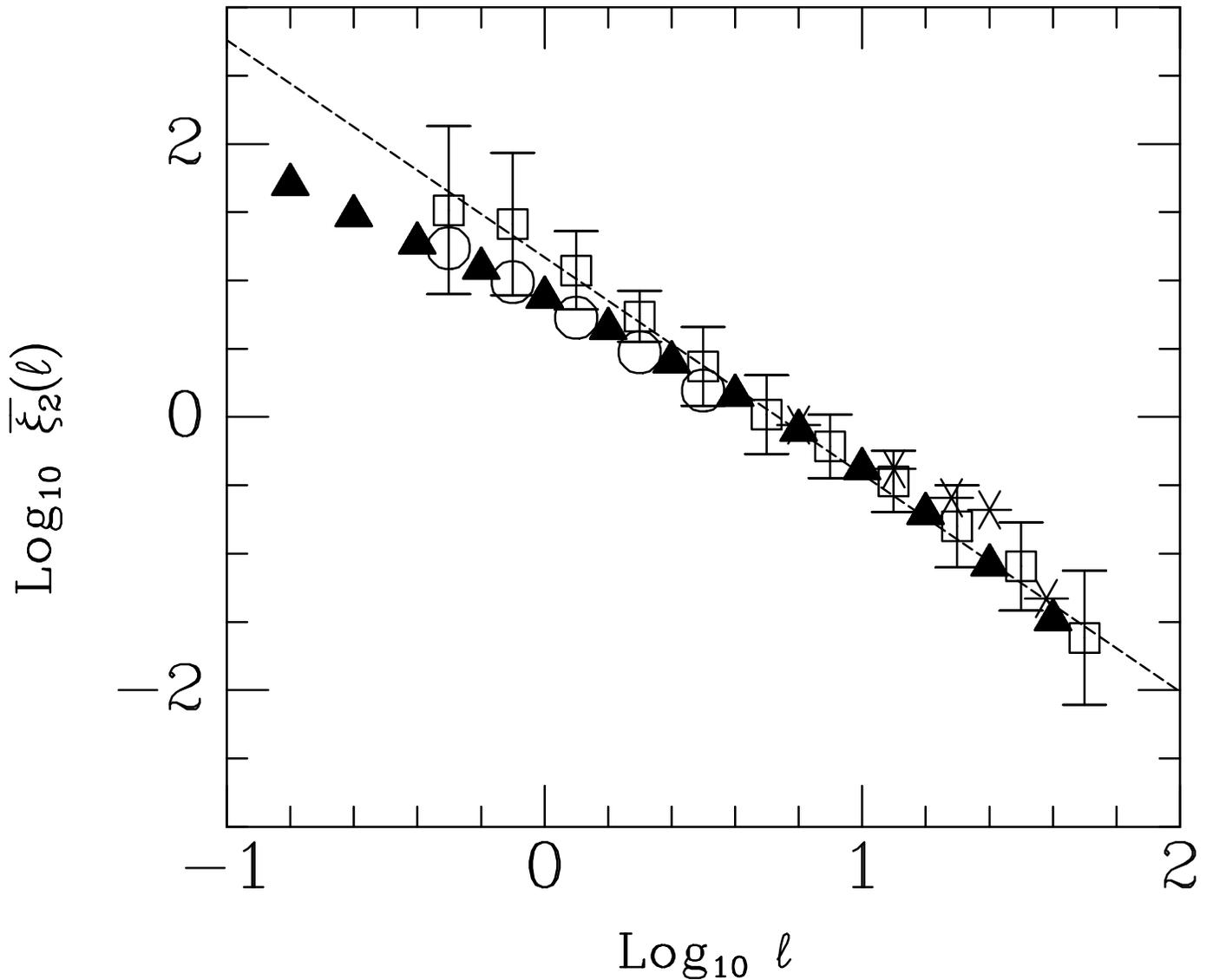

Figure 3: Average two-body correlation function $\overline{\xi_2}(\ell)$ obtained by making an equal weight average (open squares) of the raw determinations of Fig. 2. The solid triangles show a numerical integration (Eq. 2) of the best direct determination of $\xi_2$ obtained by Fisher *et al.* (1993b). The open circles show the correlation function $\overline{\xi_2}$ derived from the 2400 km s$^{-1}$ subsample alone; it is in good agreement with the triangles, and the discrepancy between the circles and squares is a reflection of the variation in the strength of the clustering as a function of luminosity. Also plotted (stars) are the (scaled) results from the measurements in the QDOT sample by Efstathiou *et al.* (1990), as well as a least square fit to the data in squares (short dashes).

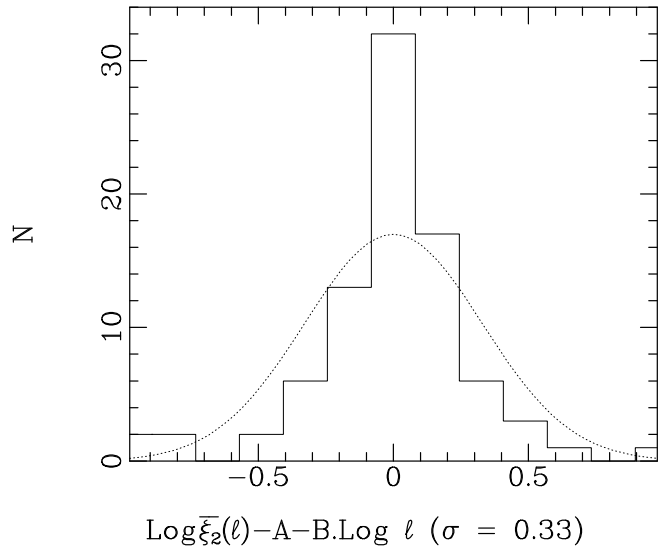

Figure 4: Distribution of deviations, summed over all scales, of the raw determinations of $\overline{\xi_2}(\ell)$ (plotted in Fig. 2) from our least squares fit to it (short dashes of Fig. 3). Also plotted for comparison is a Gaussian distribution (dots) with the same standard deviation as that of the histogram, $\sigma = 0.33$.

Fisher *et al.* (1993a; 1993b) show that the power spectrum and correlation function of *IRAS* galaxies are both in agreement with the angular correlation function derived from the APM survey (Maddox *et al.* 1990). The agreement on large scales between the Fisher *et al.* (1993b) volume-averaged correlation function, and that derived in this paper, thus implies an agreement between the present results and those of the APM.

Efstathiou *et al.* (1990) measured the variance in counts in cubical cells, $\overline{\xi_2}^c$, in a sparser and deeper redshift survey (Rowan-Robinson *et al.* 1990; hereafter QDOT) of *IRAS* galaxies. In order to compare their measurements with ours, we used the relation given by Saunders *et al.* (1991), $\overline{\xi_2}(0.63\,\ell) \simeq \overline{\xi_2}^c(\ell)$, which holds both for white noise and for a Gaussian random field with a power law correlation function of index $-1.6$. Their results at large scale are shown as stars in the figure. Although they are consistent with our results, they are on the high side. It seems from their Fig. 1 that this might be attributed to counts in a single shell; thus the discrepancy is associated with just a few cells. Fisher *et al.* (1993a) come to a similar conclusion through a power spectrum analysis of the present data set.

We now turn to higher order correlation functions, shown in Fig. 5. The top panel shows $\overline{\xi_3}^{1/2}$, the middle panel shows $\overline{\xi_4}^{1/3}$, and the bottom panel shows $\overline{\xi_5}^{1/4}$; we take roots of the data with the scale-invariant predictions in mind (Eq. 7). The results from all subsamples are shown, but in practice, the higher-order correlations are not determined from the sparsest sub-samples. The results of least square fits to the data are summarized in Table 2. Again, we see systematic effects as a function of sample volume, somewhat more pronounced than for $\xi_2$. In particular, the correlation strength on small scales is larger for more luminous galaxies. At large scale, the correlations steepen, as expected from finite volume effects. The determination of $\overline{\xi_5}$ is quite noisy, with little dynamic range in any one subsample. Nevertheless, we see that it is positive even at rather large scales.

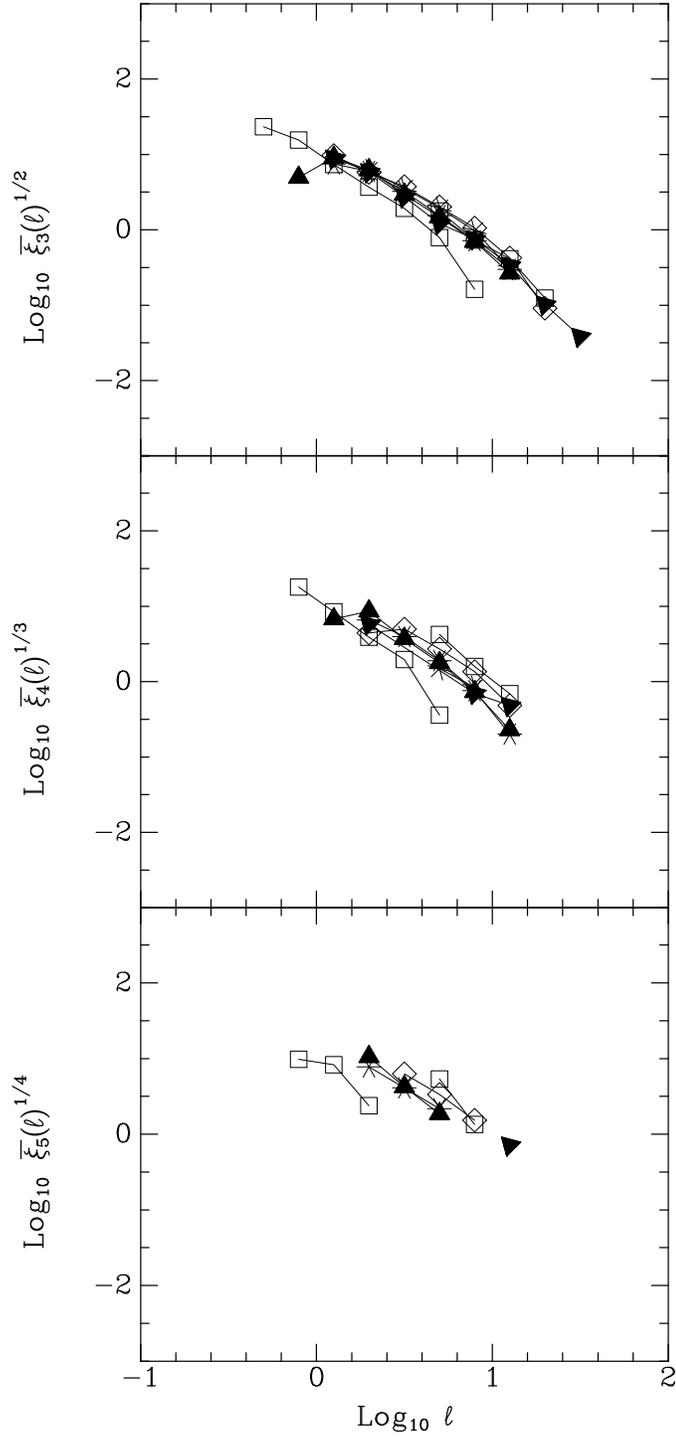

Figure 5: Raw determinations of the $N$-body correlation function, $\overline{\xi_N}^{1/(N-1)}(\ell)$, for all 10 volume-limited subsamples. The top panel corresponds to $N = 3$, the middle panel to $N = 4$, and the bottom panel to $N = 5$. The higher-order correlations are not determined from the sparsest subsamples, which is why fewer than ten curves are apparent in these plots.

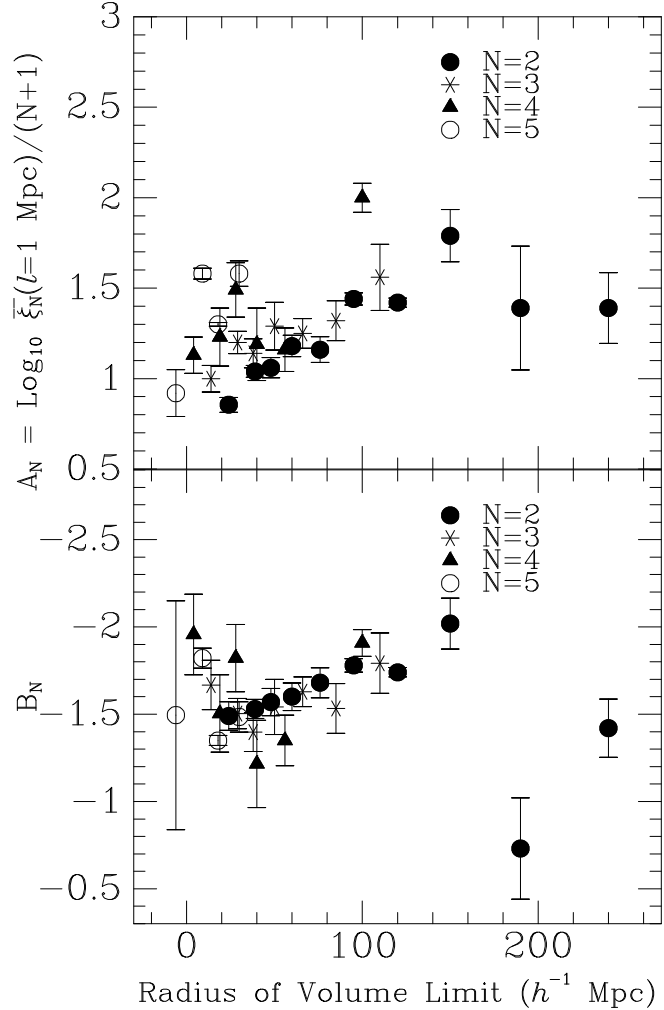

Figure 6: Results of least-squares fits of power laws to the correlation functions as a function of the subsample size, for each subsample separately. Thus the data are fit to the form $1/(N-1)\log_{10}\overline{\xi_N} = A_N + B_N \log_{10}\ell$. The results for different $N$ are given different symbols, and are staggered slightly in the ordinate to avoid overlapping error bars. These are the data tabulated in Table 2.

These systematic effects are quantified in Fig. 6, which shows the variations of the best fit coefficients $A_N$ and $B_N$ to the form $1/(N-1)\log_{10}\overline{\xi_N} = A_N + B_N \log_{10}\ell$, for $N = 2, 3, 4,$ and 5, as a function of the size of the subsample. These are the data tabulated in Table 2. The quantity $A_N$ is the logarithm of the correlation function at $\ell = 1 h^{-1}$ Mpc. This figure shows that there is an increase of clustering strength on small scales with luminosity for all correlations examined, as we saw directly from Figs. 2 and 5; for sample sizes above $100 h^{-1}$ Mpc, the data are too noisy to check if the trends continue. There is also a trend that the more luminous galaxies show a steeper slope. In the next section, we will show that these luminosity effects do *not* affect the comparison of the higher-order $\overline{\xi_N}$'s with $\overline{\xi_2}$.

Fig. 7 shows the correlation function averaged over subsamples of Fig. 5, both with clusters collapsed (open circles) and without, which has been our default (closed circles). First let us concentrate on the closed circles alone. After scaling by $(N-1)$, the $N$-point correlation functions are in very close agreement, as the scale-invariant hypothesis would predict (Eq. 7). However, the higher-order correlations are noticeably steeper than are the lower orders, which is in the sense expected from finite volume effects (Colombi *et al.* 1993a). We will quantify the comparison between the different orders in the next section. Collapsing clusters increases all the correlations on small scales, as is expected, and the effect is stronger for the higher-order correlations. Thus the higher-order correlation functions are steeper in the cluster-collapsed case than when clusters are not collapsed. Reality lies somewhere between these two extremes; the Finger of God in a cluster dilutes the true clustering on small scales, while collapsing all cluster galaxies to a single point exaggerates it. However, the correlations on scales above $3h^{-1}$ Mpc are insensitive to this, giving us greater faith in the large-scale correlations seen, even at fifth order.

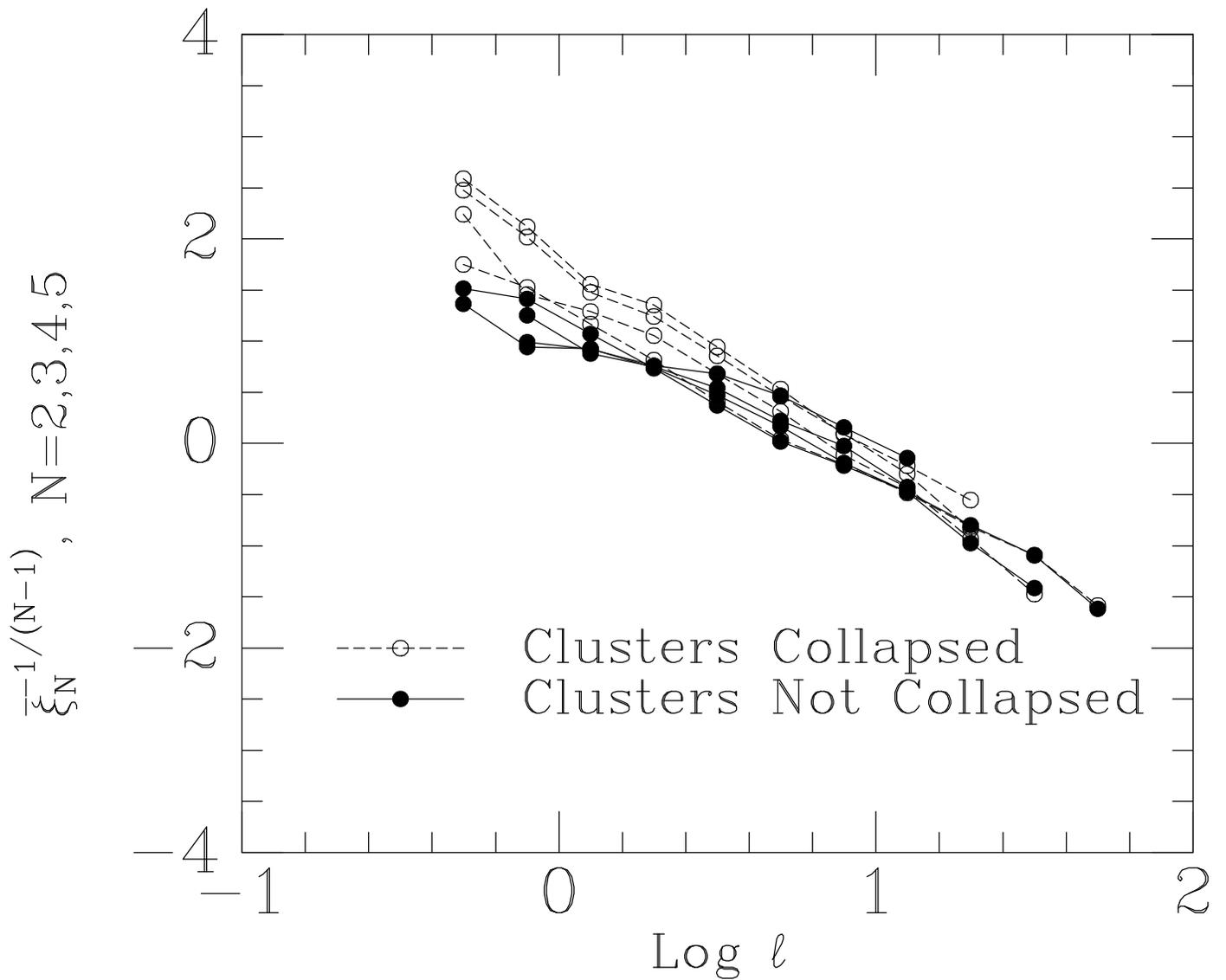

Figure 7: The 2, 3, 4, and 5-point correlation functions, averaged over subsamples, for the cases in which clusters are not collapsed (the default in this paper; closed symbols) and when they are collapsed (open symbols). The collapsing process boosts the correlations on small scales.

*3.2. Relations between the correlation functions*

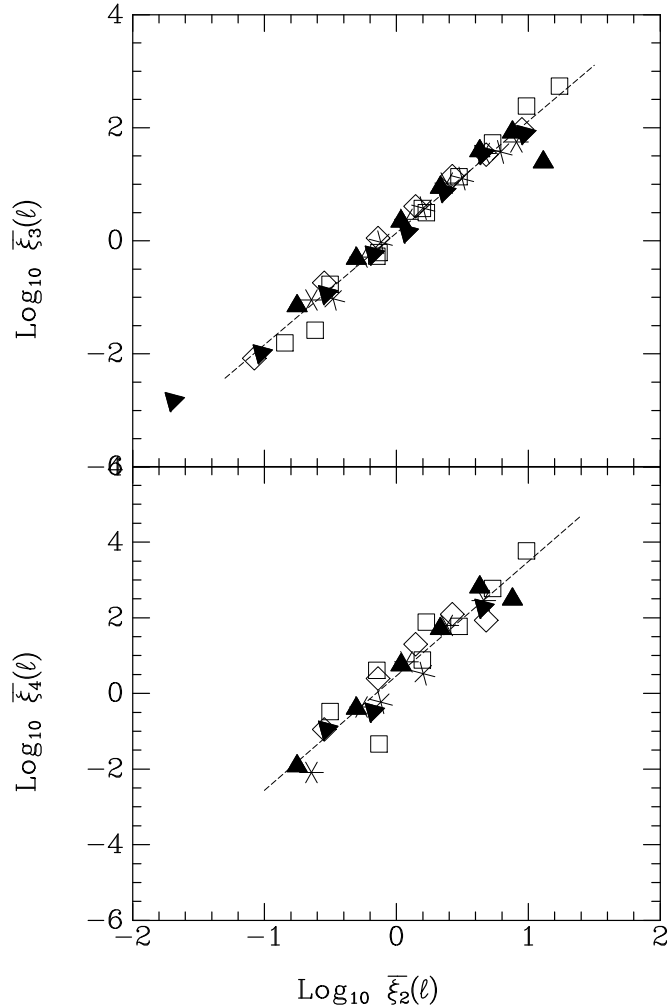

Figure 8: Determinations of $\overline{\xi_N}$ vs. $\overline{\xi_2}$ in our series of volume-limited sub-samples ($N = 3$, top panel, and $N = 4$, bottom panel). The dashed lines show least squares fits to the points.

We just saw that the correlation functions are well-described by power laws, and the least square fits we made suggest that the high order $\overline{\xi_N}$ can be expressed simply as powers of $\overline{\xi_2}$. Fig. 8 checks this directly by plotting $\overline{\xi_3}$ and $\overline{\xi_4}$ against $\overline{\xi_2}$. The relations are remarkably tight. Least square fits to the points in the figures, $\log_{10} \overline{\xi_N} = C_N + D_N \log_{10} \overline{\xi_2}$, yield $C_3 = 0.15 \pm 0.05$ and $D_3 = 1.96 \pm 0.06$, and $C_4 = 0.46 \pm 0.09$ and $D_4 = 3.03 \pm 0.18$ (the point of lowest $\overline{\xi_2}$ was not taken into account at the third order). One worry is the luminosity dependence of the strength of the correlations on small scales which we saw above. Table 2 and Fig. 9 show the results of least-square fits to the $\overline{\xi_N} - \overline{\xi_2}$ data for each subsample independently; no systematic effects are seen. That is, within the noise, any luminosity effects in the correlations are such as to move points along the lines in Fig. 8, not perpendicular to them. In other words, the fact that galaxies of different luminosities show slightly different levels of bias does not grossly affect the relationships between the moments. The horizontal lines in the lower panel of the figure show the scale-invariant predictions; the fits to the individual subsamples indeed do cluster around them, although the scatter is of course worse as $N$ increases. Irrespective of any theoretical prejudice, the data does indicate rather unequivocally that the skewness and the kurtosis are indeed proportional to (respectively) the square and the cube of the variance over the whole range accessible. Interestingly, the derived $D_N$ do not agree as well with the scale-invariance predictions in the collapsed clusters case (cf., Fig. 7), and the discrepancy increases with order.

Given the previous results, we attempt to go one step further and plot in Fig. 10 the ratio $S_3 \equiv \overline{\xi_3}/\overline{\xi_2}^2$ versus the variance. The upper panel shows the raw determinations for each subsample; average values are given in Table 2. In the lower panel, the open squares show the mean value of $S_3$ in bins of values of $\log_{10} \overline{\xi_2}$; the plotted error bars correspond to one standard deviation in each bin. There is a weak trend of increasing $S_3$ with variance, but the data are also compatible with a constant value of $S_3$, even from the quasi-linear to the highly non-linear regimes. The average over all values is $S_3 = 1.5 \pm 0.5$, which is indicated by the dashed line. This error is larger than would be expected from the linear fit to points in Fig. 8, both because of the difference between linear and logarithmic weighting, and because two parameters are fit to the data in Fig. 8, while only one is fit here. Note that Table 2 shows that a least-square fit to the correlation function gives different slopes for $\overline{\xi_2}(\ell)$ and $(\overline{\xi_3}(\ell))^{1/2}$, implying that $S_3$ cannot be perfectly constant. However, the data are noisy, and this difference in slope (and thus the variation in $S_3$) is not significant; after all, we found that $D_3$ was consistent with the scale-invariant prediction of 2.0.

The three error bars in the weakly non-linear regime show the ratio of the skewness and variance squared measured in Gaussian windows in the QDOT sample (Saunders *et al.* 1991; cf., Coles & Frenk 1991). Since these latter authors did not look directly at $S_3$, we have taken the values they quote for the variance and the skewness and used the standard propagation of errors.

There is another approach to the measurement of $S_3$: at small scales, it is observed that the three-point correlation function may be expressed as a symmetrized sum of double products of two-point correlation functions times a constant called $Q$ (Peebles 1980). It then follows that $\overline{\xi_3} = 3Q J_3 \xi_2^2$, where $J_3$ is given by

$$J_3 \equiv (3/4\pi)^3 \int_0^1 d^3\mathbf{x}_1 d^3\mathbf{x}_2 d^3\mathbf{x}_3 (||\mathbf{x}_1 - \mathbf{x}_2|| \, ||\mathbf{x}_2 - \mathbf{x}_3||)^{-\gamma} \quad . \tag{19}$$

Thus $S_3 = 3Q J_3/J_2^2$, where $J_2$ is given by Eq. 10. For the *IRAS* galaxies, we set $\gamma = 1.59$ (see Table 2) and find $J_3 = 2.69$ and thus $S_3 = 3.17\,Q$, yielding $Q = 0.49 \pm 0.16$. Alternatively, for $\gamma = 1.77$ appropriate for optically selected galaxies, we find $J_3 = 3.69$, and so $S_3 = 3.34\,Q$. The triangles on the right of the figure correspond to the values of $Q = 1.29 \pm 0.21$ (Groth & Peebles 1977), and $Q = 0.8 \pm 0.07$ (Peebles 1980). The latter value, which yields a value of $S_3$ consistent within $1.5\,\sigma$ with that determined from the present sample, was obtained by analyzing the galaxy catalog of Davis, Geller, & Huchra (1978) which was chosen to contain no prominent clusters, and is in agreement with the value measured from the angular distribution of *IRAS* galaxies by Meiksin *et al.* (1992). This is the expected trend, since the skewness measures the left-right asymmetry of the distribution, *i.e.*, the degree of asymmetry between overdense and underdense regions. The lower value of $S_3$ for *IRAS* galaxies than for optically selected galaxies just reflects the under-representation of *IRAS* galaxies in dense cluster cores.

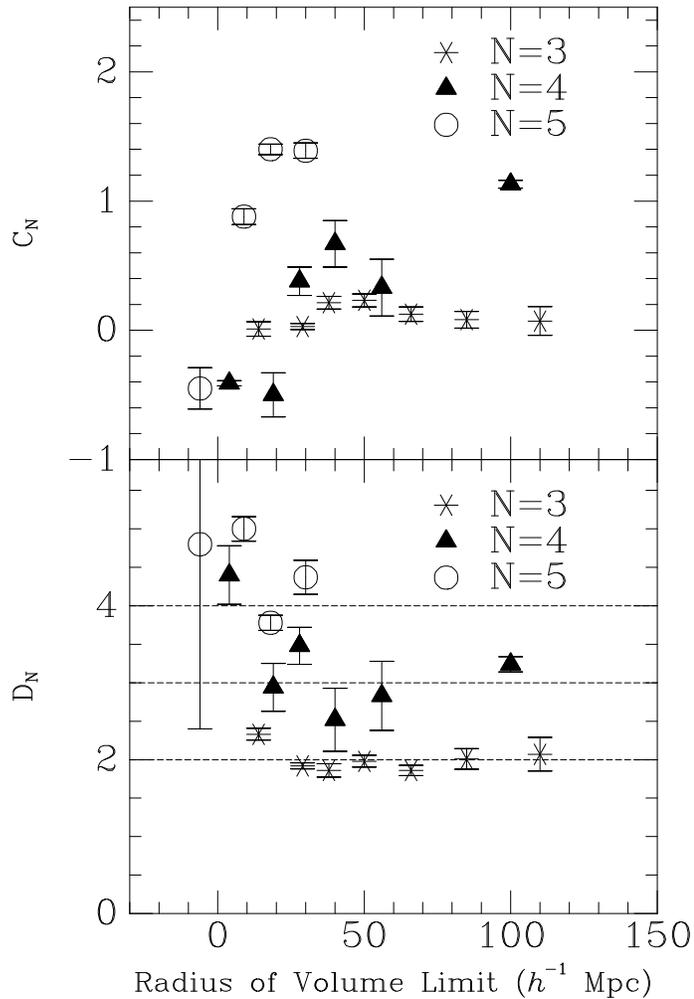

Figure 9: Results of least-squares fits of power laws to the correlation functions $\overline{\xi_N}$ as a function of $\overline{\xi_2}$, for each subsample separately. Thus the data are fit to the form $1/(N-1)\log_{10}\overline{\xi_N} = C_N + D_N\overline{\xi_2}$. The results for different $N$ are given different symbols, and are staggered slightly in the ordinate to avoid overlapping error bars. These are the data tabulated in Table 2. In the lower panel, the dashed lines indicate the predictions of the scale-invariant hypothesis: $D_3 = 2$, $D_4 = 3$, and $D_5 = 4$.

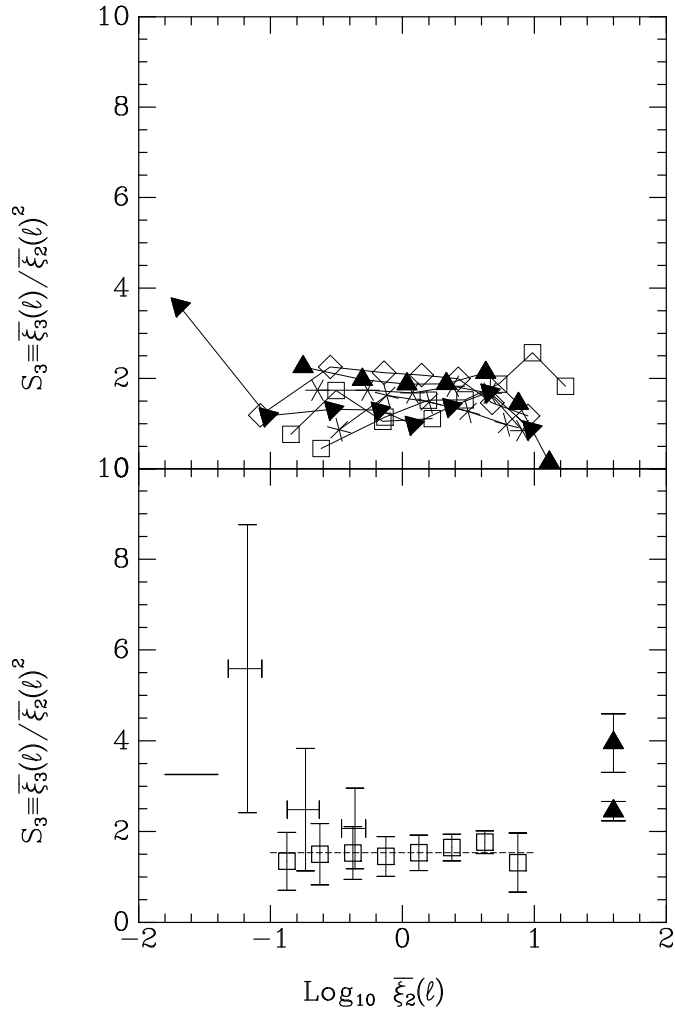

Figure 10: All determinations of $S_3 \equiv \overline{\xi_3}/\overline{\xi_2}^2$ in our series of volume-limited sub-samples are presented in the top panel. The bottom panel shows an equal weight average in bins of values of $\log_{10} \overline{\xi_2}$, the average value (dashes), as well as the values inferred from measurements of $Q$ in the non-linear regime from optical data (triangles), from measurements of skewness and variance on the QDOT sample (error bars on left) and the theoretical prediction (solid line on left) from perturbation theory for a power spectrum of index $n = -1.4$ and no bias ($b = 1$).

Juszkiewicz *et al.* (1993a) use perturbation theory to calculate the value of $S_3$ resulting from number counts in spherical cells, and find $S_3 = 34/7 - (n + 3)$, where $n$ is the index of the power spectrum ($P(k) \propto k^n$). The solid line on the left of the figure indicates the value expected for $n = -1.4$ (and $b = 1$, see below), which is appropriate for the scales probed by our measurements (Fisher *et al.* 1993a). Similar predictions for $S_3$ can be made in the case of a Gaussian smoothing appropriate to the QDOT measurements (cf., Juszkiewicz *et al.* 1993a).

As the results for the skewness depend on the biasing model relating the underlying matter density field $\delta_M$ to that traced by the galaxies $\delta$, our results put constraints on the model. For simple linear biasing in which

$$\delta = b \, \delta_M \quad , \tag{20}$$

we have

$$S_3 = \left\langle (b \, \delta_M)^3 \right\rangle / \left\langle (b \, \delta_M)^2 \right\rangle^2 = S_{3M}/b \quad , \tag{21}$$

where $S_{3M}$ is the value appropriate for the dark matter, as given by the horizontal line on the left of the figure. Thus, for $n = -1.4$, our determination of $S_3$ from the data yields $1.6 < b < 3.2$ ($1\,\sigma$). While this is the range of values currently in fashion for the bias parameter (e.g., Weinberg 1989), this result should be taken with a large grain of salt; the value of the bias parameter refers to a model in which positive and negative densities are treated in equivalent ways (Eq. 20), while the derivation of $b$ above is based on the *asymmetry* in the density field. Even more damning is the sensitivity of $S_3$ to the treatment of the clusters of galaxies. We will see in §3.3 that the derived value of $b$ by Eq. (21) when clusters are given extra weight to match the overdensities seen in optically selected samples of galaxies (Strauss *et al.* 1992a), is less than half the value derived here, and thus this estimate of $b$ has a systematic uncertainty of at least a factor of two. Finally, any non-linear term in the biasing scheme modifies the relation between $S_3$ and $S_{3M}$ given in Eq. 21, although it does preserve the constancy of $S_3$ in the weakly non-linear regime (Fry & Gaztañaga 1993; Juszkiewicz *et al.* 1993b).

Fig. 11 is the exact analog of Fig. 10 for the fourth central moment (kurtosis) of the counts distribution. The data are given in Table 2. Although the dispersion is much larger than in the skewness case, we do find that $S_4$ is approximately constant (within the error bars) over the whole range accessible. If we assume that $S_4$ is a constant, we find $S_4 = 4.4 \pm 3.7$. Even though this value is rather ill-determined, and one cannot exclude the possibility of a substantial trend of $S_4$ with scale (or variance), one must bear in mind the 6 orders of magnitude spanned by $\overline{\xi_4}$ and $\overline{\xi_2}^3$.

We stop here at the fourth moment. However, we saw in Fig. 5 above that we were able to derive the fifth moment from the data, and that qualitatively at least it obeyed the scaling laws described in the Introduction. However, the slopes of the best-fit lines to the second-order and fifth order correlation functions are not in perfect agreement, and we saw above that noise in the determination of $S_3$ and $S_4$ did not allow us to conclude definitively that they were independent of scale. These problems only get worse at higher moments (cf., Table 2 and Fig. 9). Moreover, as discussed at the beginning of § 4, the higher-order moments become increasingly sensitive to the very rare peaks which may be completely missed in a finite volume. Fortunately, there are ways to check whether scale invariance holds at high orders which do not rely on the direct determination of the correlation functions, which we explore in §4.

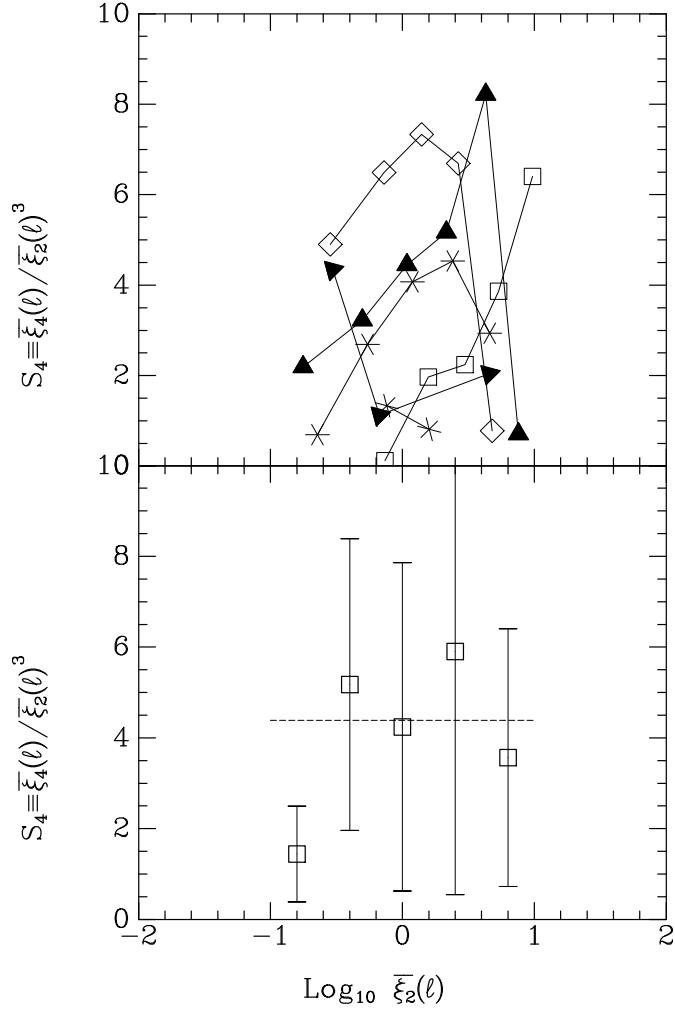

Figure 11: All determinations of $S_4 \equiv \overline{\xi_4}/\overline{\xi_2}^3$ in our series of volume-limited sub-samples are presented in the top panel. The bottom panel shows an equal weight average in bins of values of $\log_{10} \overline{\xi_2}$, as well as the average value $S_4 = 4.4$ (dashes).

*3.3. The effect of boosting clusters*

Strauss *et al.* (1992a) show that *IRAS* galaxies give systematically lower density estimates in the cores of clusters than do optically selected galaxies. The higher-order correlation functions depend sensitively on the tails of the density distribution, which motivates us to test the sensitivity of our results to the cluster densities. Strauss *et al.* (1992a), in their Table 2, derive density estimates in the central 100 km s$^{-1}$ of the cores of seven nearby clusters; we have performed counts on the *IRAS* sample, in which galaxies associated with the clusters are given extra weight corresponding to the ratio of the optical and *IRAS* density estimates. This affects a total of only 126 galaxies (not all of which enter into one of the volume-limited subsamples). Nevertheless, the results are dramatic. As expected, all correlations are strengthened, especially for larger $N$. We first show the relations between the $\overline{\xi_N}$ in Fig. 12. The fit to a line is even better than it was before, and extends over a greater range of variance (note the difference in scales from Fig. 8). The least-square fits to the points are shown in the figure, and give $C_3 = 0.56 \pm 0.02$, $D_3 = 2.06 \pm 0.01$, and $C_4 = 1.30 \pm 0.04$, $D_4 = 3.18 \pm 0.03$. The slopes here are now slightly steeper than before, and are no longer consistent, within the errors, with the scale-invariant prediction. This is reflected in the derived values of $S_3$, shown in Fig. 13. The relative constancy of $S_3$ seen in Fig. 10 is no longer seen; in particular, $S_3$ is an increasing function of $\overline{\xi_2}$. In addition, the average value of $S_3$ is appreciably higher in the present case. The lower panel shows an equal weight average in bins of $\log_{10} \overline{\xi_2}$. The dashed line is the average of the points with $0.1 < \overline{\xi_2} < 10$, namely $3.71 \pm 0.95$, which is more than twice the value we found for the unboosted case (cf., Fig. 10). This emphasizes the danger of deriving a value of the bias $b$ from these data; by double-counting $\sim 2\%$ of the galaxies, we decrease the bias estimate by more than a factor of two, to $0.70 < b < 1.18$. Note that *boosting* clusters (and thus increasing the correlations) causes the bias estimate to go *down*; this is because the three-point correlation function is affected even more than the two-point correlation function. Of course, this is a very non-linear transformation of the density field; as we mentioned above, Fry & Gaztañaga (1993) and Juszkiewicz *et al.* (1993b) show that non-linear transformations add extra terms to Eq. 21. Not surprisingly, $S_4$ is changed even more than $S_3$ by the boosting; we find $S_4 = 23.6 \pm 12.1$ for the range $0.1 < \overline{\xi_2} < 10$ (compare with the value $4.4 \pm 3.7$ found above), with even a more dramatic rise in the densest regions.

## 4. Void probability $P_0$

The counts in cells also provide measurements of the void probability function $P_0$. As discussed in the Introduction, the deviations of $P_0$ from its Poisson value can be quantified by the function $\sigma$; under the scale-invariance hypothesis, $\sigma$ depends only on $N_c$ (Eqs. 1 and 7). The behavior of $\sigma(n, v)$ thus provides us with an indirect way of testing the scaling of high order moments with variance.

The top panel of Fig. 14 shows the measurements of $\sigma \equiv -\log P_0/\overline{N}$ versus $\ell$ in our series of volume-limited samples (which of course are of varying number density), while the bottom panel shows the same measurements as a function of the number of clustered particles above the mean $N_c(\ell)$. It clearly demonstrates that $\sigma$ behaves according to the prediction of scale-invariant models to an amazing degree of accuracy. The agreement between the determinations in different (nearly independent) subsamples also suggests that the error bars in each of them are rather small, typically of the size of the plotting symbols themselves, for all but the point of largest $N_c$ in each subsample.

How might we make a quantitative assessment of the errors in $\sigma$? By comparing numerical simulation results and analytical calculations, Colombi *et al.* (1993b) obtain an upper limit for the error

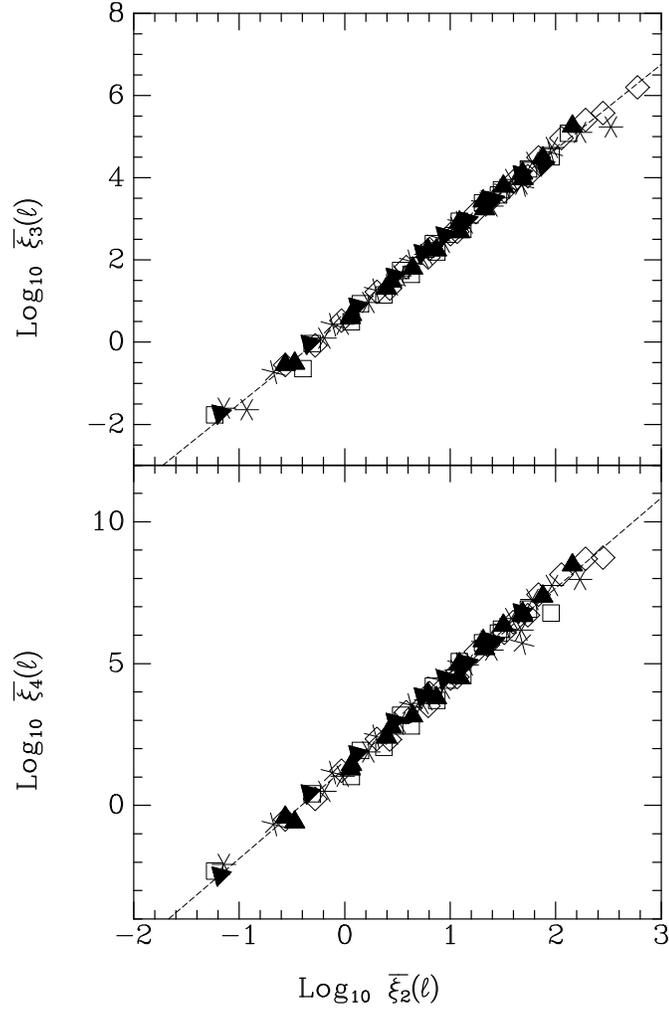

Figure 12: Determinations of $\overline{\xi_N}$ vs. $\overline{\xi_2}$, as in Fig. 8, now with cluster galaxies boosted. Note that the scale has changed from Fig. 8. The dashed lines show least squares fits to the points.

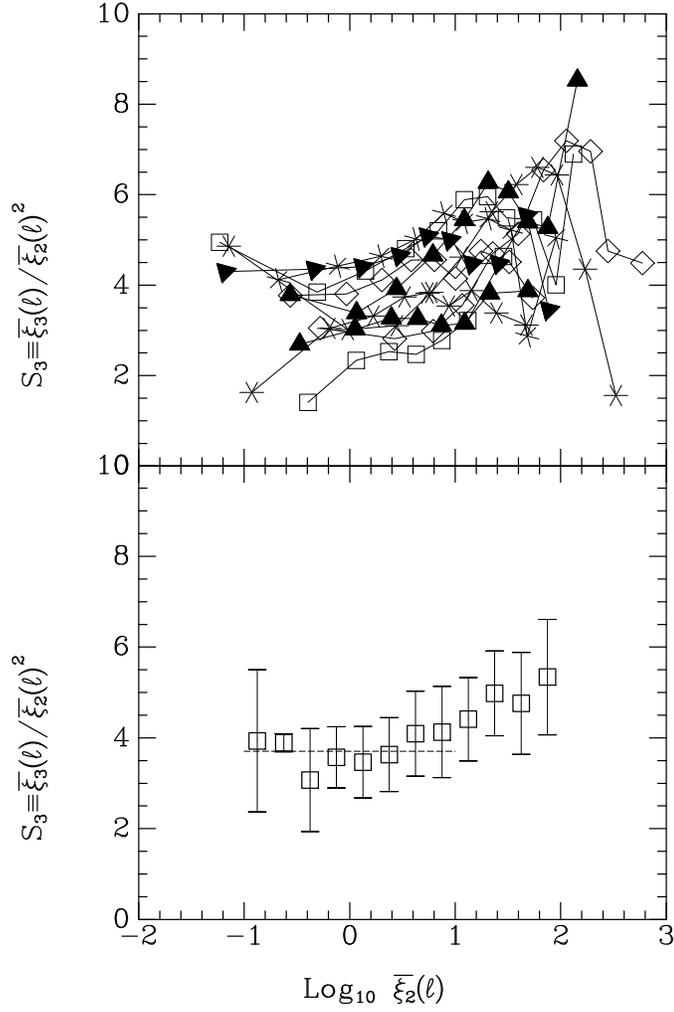

Figure 13: All determinations of $S_3 \equiv \overline{\xi_3}/\overline{\xi_2}^2$ from the boosted counts are presented in the top panel. The bottom panel shows an equal weight average in bins of values of $\log_{10} \overline{\xi_2}$ and the average value (dashes).

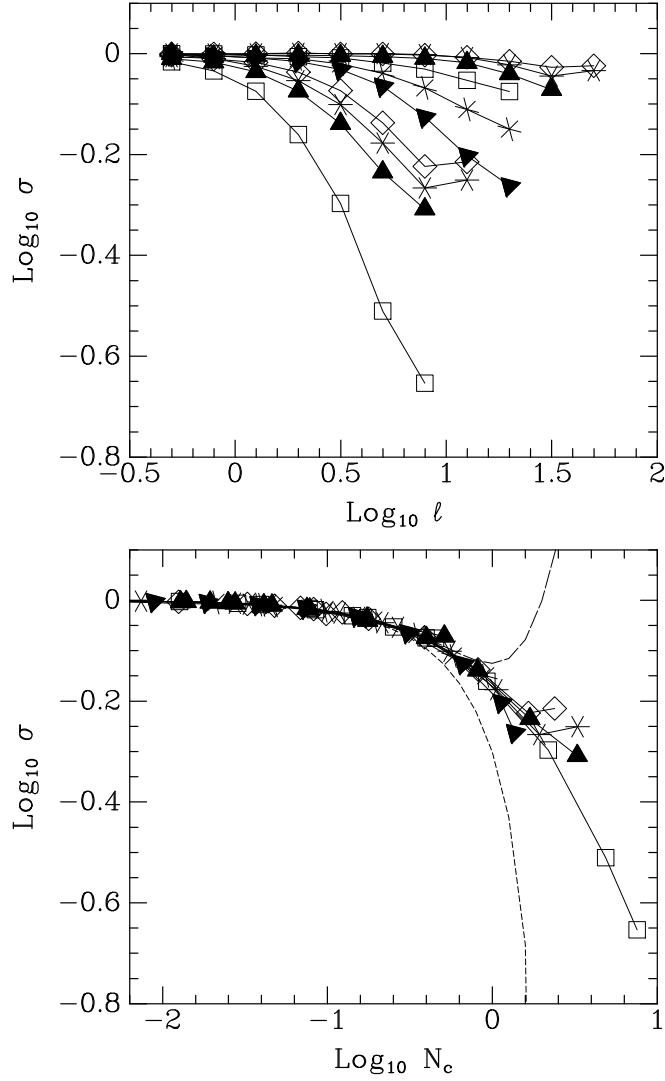

Figure 14: Departures from Poisson statistics $\sigma = -\log P_0/\overline{N}$, plotted in the top panel as a function of the cell sizes $\ell$ : from bottom to top the curves correspond to volume-limited samples of increasing volume. In the bottom panel, $\sigma$ is expressed as a function of $N_c(\ell)$. The short dashed curve is the prediction $\sigma = 1 - N_c/2$ expected if correlations higher than second order are negligible, while the long dashed curve is the prediction $\sigma = 1 - N_c/2 + S_3 N_c^2/6$ expected if correlations higher than third order are negligible.

in $\sigma$ in a volume of size $L_{sample}$:

$$\frac{\Delta\sigma}{\sigma} \lesssim \left(\frac{1}{N_{tot}} + \overline{\xi}_2(L_{sample})\right)^{1/2} \quad , \tag{22}$$

where $N_{tot} = \overline{N}(4\pi/3)L_{sample}^3$ is the total number of galaxies in the volume, and $\overline{\xi}_2(L_{sample})$ is the sample-to-sample variance. Eq. (22) is valid for values of $\ell$ such that there is less than one void of size $\ell$ in the volume; that is, $P_0(4\pi/3)L_{sample}^3 \ell^{-3} \lesssim 1$, a condition violated for the point of smallest $\ell$ (largest $N_c$) for each subsample in Fig. 14. Eq. (22) estimates the true error in $\sigma$ by a large amount when $\sigma \ll 1$. Of course, the abscissa in Fig. 14 is also subject to error; an estimate of the fractional uncertainty in $N_c$ is $(S_4/2\overline{\xi}_2(L_{sample}))^{1/2}$. For the five smallest volumes, the shot noise contribution to the error in $\sigma$ in Eq. (22) is negligible, and with $S_4 \approx 4$ we find: $\Delta\sigma/\sigma \simeq \Delta N_c/N_c \lesssim \overline{\xi}_2^{-1/2} \simeq 0.12, 0.08, 0.07, 0.06$, and $0.05$ respectively with increasing sample volume. Note that the ordinate in Fig. 14 is $\log_{10}\sigma$; thus a fractional error in $\sigma$ of 0.12 corresponds to an error bar in the figure of 0.05.

It is important to note that the regime tested corresponds to relatively small values of $N_c(\ell)$, i.e., $N_c(\ell) \lesssim 1$, and we are thus only probing the *dilute* regime ($\sigma \approx 1$), where departures from Poisson are rather small. Still, if all correlations beyond second order are negligible, then Eq. 1 shows that $\sigma = 1 - N_c/2$ (short dashes in the figure); it is clearly a bad fit. If we can ignore terms beyond third order, then $\sigma = 1 - N_c/2 + S_3 N_c^2/6$, which is plotted as the long dashes in the figure, using $S_3 = 1.5$, as we found above. This also is a poor fit, indicating that higher-order correlations are not negligible. This clearly shows that, when $N_c \sim 1$, one probes the non-perturbative regime where *all* orders contribute.

Nevertheless, the various scale-invariant models that have been proposed for the specific ensemble of values for the $S_N$ differ substantially from one another only in the regime $N_c \gg 1$, and become degenerate in the sparse sampling limit (Colombi *et al.* 1993a). Balian & Schaeffer (1989a) argue that scale-invariance implies that $\sigma \propto N_c^\omega$ for $N_c \gg 1$, where $\omega$ is a constant which characterizes the non-linear clustering of the sample. Fig. 14 shows only marginal evidence that $\sigma$ asymptotically approaches a power-law; only the densest sample we have, volume-limited to 2400 km s$^{-1}$, probes the regime $N_c > 1$.

Various models have been proposed for the clustering hierarchy, which we are now in a position to test directly. Among them are the negative binomial model (Carruthers & Shih 1983) which provides a good fit to the CfA data (Gaztañaga & Yokohama 1993):

$$P_N = \frac{1}{N!}\overline{N}^N(1+\overline{N}\,\overline{\xi}_2)^{-N-1/\overline{\xi}_2}\prod_{j=1}^{N-1}(1+j\overline{\xi}_2) \quad . \tag{23}$$

In this model,

$$\sigma \equiv -\log P_0/\overline{N} = \log(1+N_c)/N_c \tag{24}$$

is a function of $N_c$ alone, making it consistent with the scale-invariant hypothesis. Saslaw & Hamilton (1984, see also Saslaw 1985; 1989) have used thermodynamic arguments to propose that the counts in cells are given by

$$P_N = \frac{\overline{N}(1-b)}{N!}\left[\overline{N}(1-b)+Nb\right]^{N-1}\exp\left[-\overline{N}(1-b)-Nb\right] \quad , \tag{25}$$

where $b = 1 - 1/(1+N_c)^{1/2}$, if the model is to have the same variance as the data (Fry 1986). In this case,

$$\sigma = 1/(1+N_c)^{1/2} \quad , \tag{26}$$

and this model is also consistent with the scale-invariant hypothesis.

Coles & Jones (1991) hypothesize that the density field is given by a log-normal distribution, which implies that the counts in cells are given by:

$$P_N = \frac{1}{(2\pi)^{1/2}\sigma} \frac{1}{N!} \int_0^\infty \lambda^{N-1} e^{-\lambda} \exp\left[-\frac{(\log \lambda - \log \alpha)^2}{2\sigma^2}\right] d\lambda \quad , \tag{27}$$

where

$$\alpha = \frac{\overline{N}}{(1+\overline{\xi_2})^{1/2}} \text{ and } \sigma = (\log(1+\overline{\xi_2}))^{1/2} \quad . \tag{28}$$

The expression $\sigma \equiv -\log P_0/\overline{N}$ is manifestly not a function of $N_c$ alone, and thus this model is explicitly non-scale-invariant.

Finally, two models have been proposed for the $P_0$ alone. Fry (1986) has developed a hierarchical Poisson model in which

$$\sigma = (1 - e^{-N_c})/N_c \quad , \tag{29}$$

and Alimi et al. (1990) use a phenomenological fit to the quantity $\sigma$ that asymptotes to a power-law at large $N_c$:

$$\sigma = (1 + N_c/2\omega)^{-\omega} \quad . \tag{30}$$

Fig. 15 shows the $\sigma - N_c$ comparison for the 2400 km s$^{-1}$ subsample, together with four models: the hierarchical Poisson model (Eq. 29, short dashes), the negative binomial model (Eq. 24, long dashes), the thermodynamic model (Eq. 26, long dot-dashes), and the phenomenological fit of Eq. 30, with $\omega = 0.9$ (short dot-dashes). The log-normal model, Eq. 27, is not included in this figure, as it fails to predict the scale-invariance of $\sigma$. The best fit is that of Eq. 30, but of course, it is the only one with a free parameter. Moreover, the largest discrepancies between the models and the data occur for values of $N_c$ which only the sample volume-limited to 2400 km s$^{-1}$ probes (compare Fig. 15 to Fig. 14). In fact, as typical errors in $\log_{10} \sigma$ are 0.05, the discrepancy between the data and the hierarchical Poisson and negative binomial models is only at the 1 $\sigma$ level. Incidentally, Alimi et al. (1990) found a best-fit value of $\omega = 0.5 \pm 0.15$ from the CfA data they analyzed, but they were able to fit to the region within which $\sigma$ asymptotes to a power-law of $N_c$, which IRAS does not probe, thus the difference between these two values is not very significant.

As discussed in the Introduction, Balian & Schaeffer (1989a) show that the scale-invariant hypothesis implies that the $P_N$ follow various specific scaling laws in various regimes. These scaling laws become manifest in the limit of high sampling, and provide the most powerful and direct tests of scale invariance. Unfortunately, the IRAS sample is simply too sparse to allow us to test these forms.

This can be illustrated as follows. The scaling laws are determined by three characteristic numbers, $\overline{N}(\ell)$, the average number of galaxies in a volume, $N_c(\ell)$, which as we saw characterizes the number of clustered particles in the volume, and $N_v(\ell) \equiv (\log P_0(\ell))^{1/(1-\omega)}$, which is equal to unity on scales $\ell$ of the characteristic sizes of voids. Interesting scaling laws appear in the limit $1 \ll N_v \ll \overline{N} \ll N_c$. Fig. 16 shows these three characteristic numbers as a function of scale for the 2400 km s$^{-1}$ subsample. The region of parameter space in which this limit is satisfied is vanishingly small, and the situation is of course worse for the larger and sparser subsamples.

With this limited ability to distinguish models in mind, we now turn to direct comparisons of the observed $P_N$ with models. Fig. 17 compares the measured $P_N$ of Fig. 1 with various models. The dots show the distribution predicted by the thermodynamic model of Saslaw (Eq. 25). Contrary to the case of numerical simulations (BSD; Bouchet & Hernquist 1992; but also see Suto, Itoh, & Inagaki 1990) this distribution provides quite a good fit to the data. The Log-Normal model (Eq. 27, short dashes) also fits the data well, as does the negative binomial model (Eq. 23, long dashes), although none properly match the steep drop-off of the $P_N$ at large $N$ in the densest sub-sample, with the log-normal distribution behaving the most poorly. However, this sub-sample is of course of the smallest volume, and finite volume effects are far from negligible (Colombi *et al.* 1993a). Thus, as in Fig. 15, we cannot use the discrepancy between the $P_N$ and the models to rule out the latter. In larger volumes, the sparseness of the samples causes all models to become degenerate. This degeneracy is broken only in the regime in which the number of clustered points $N_c$ is much larger than unity for scales smaller than the correlation length (Colombi *et al.* 1993b).

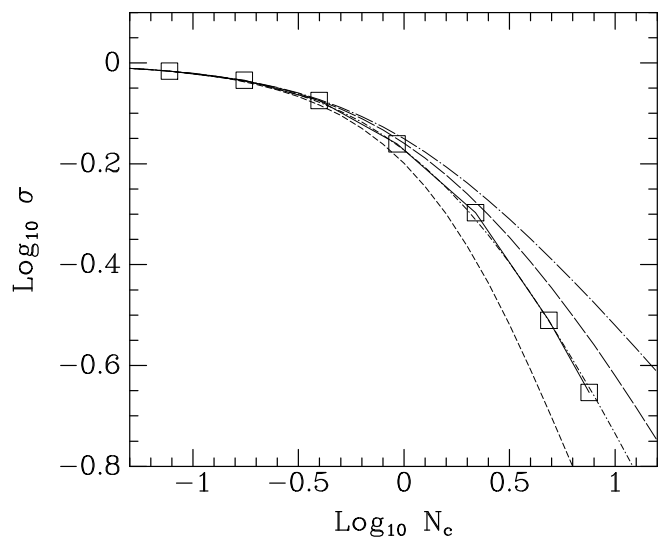

Figure 15: The quantity $\sigma$ plotted against $N_c$ for the densest subsample of Fig. 14. Also plotted are the hierarchical Poisson model of Fry (1986) (short dashes), the negative binomial model of Gaztañaga & Yokohama (1993) (long dashes), the thermodynamic model of Saslaw & Hamilton (1984) (long dot-dashes) and the phenomenological fit of Alimi et al. (1990) with $\omega = 0.9$ (short dot-dashes).

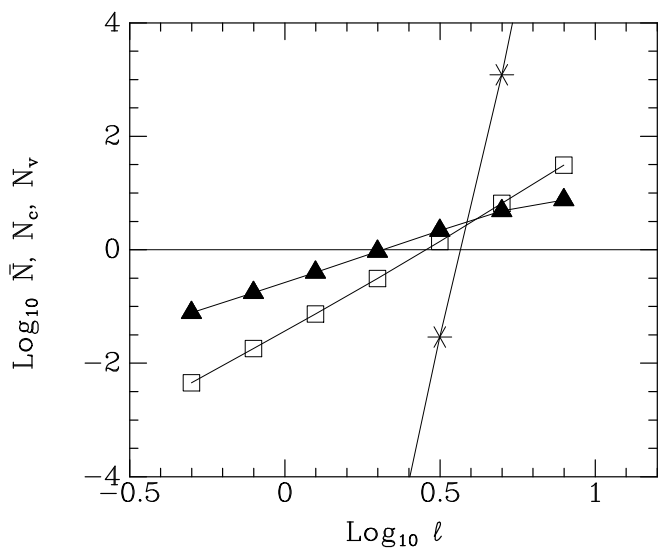

Figure 16: Variation of characteristic particle numbers, $N_c(\ell)$ (triangles), $N_v(\ell)$ (stars), and $\overline{N}$ (squares), in the densest (2400 km s$^{-1}$) sub-sample. We used $\omega = 0.9$ to derive $N_v$. A horizontal line is drawn at $N = 1$.

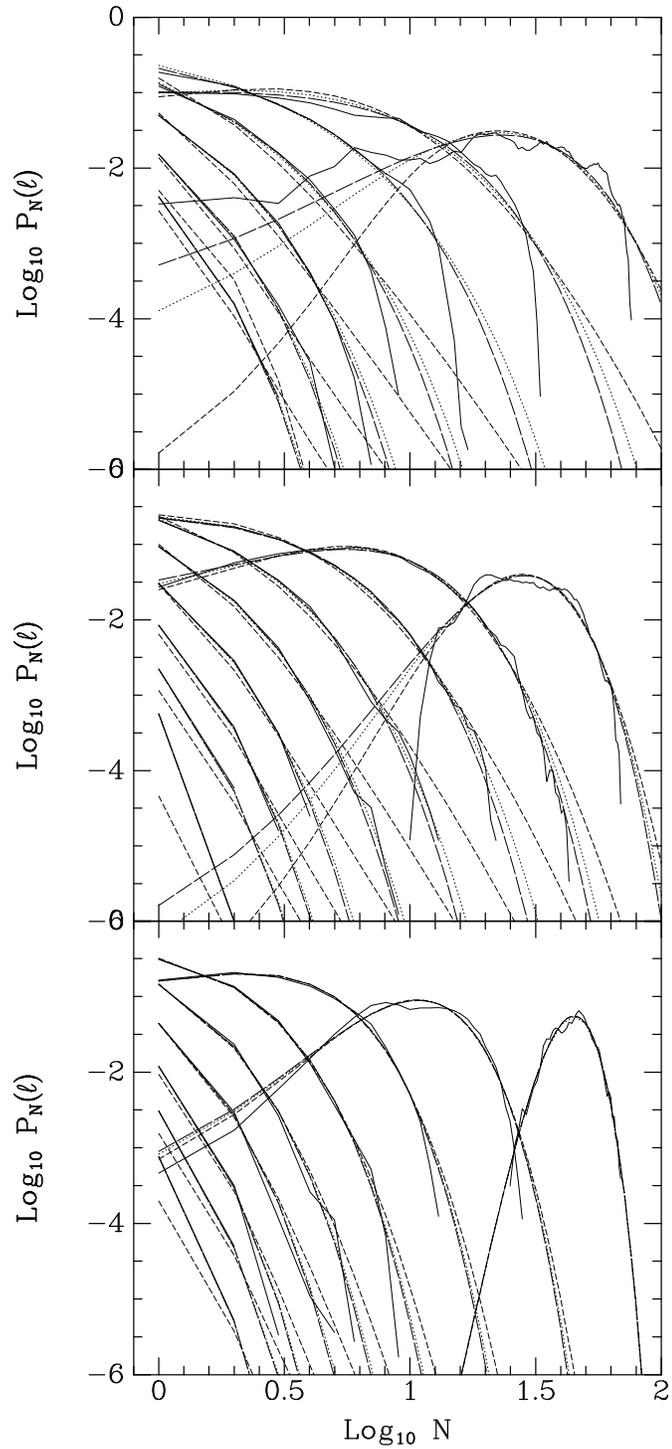

Figure 17: Same as Fig. 1, but the observed $P_N(\ell)$ are now compared with various theoretical models. The dots correspond to the distribution predicted by Saslaw once the variance is adjusted to equal that of the data. Short-dashes show a log-normal distribution, while long-dashes show a negative binomial distribution.

## 5. Conclusions

We have measured the count probability distribution in a series of 10 volume-limited subsamples, each roughly twice as big in volume as the previous one. We saw that, once the Poisson contribution is subtracted, the variance of this distribution, $\overline{\xi_2}$, is well fit by a single power law of index $\gamma = -1.59$ over 2 decades of scale (*i.e.*, for a cell *radius* $0.5h^{-1}\mathrm{Mpc} < \ell < 50h^{-1}\mathrm{Mpc}$). We have found a weak dependence of the small-scale correlation strength on *IRAS* luminosity, which appears at all orders we investigated. We derived the higher-order correlation functions, to $\overline{\xi_5}$, and found them to obey power-laws with slopes given approximately by $(N-1)\gamma$. We also found that the skewness $\overline{\xi_3}$ is closely approximated by $\overline{\xi_3} \propto \overline{\xi_2}^{1.96 \pm 0.06}$ over the range $0.1 < \overline{\xi_2} < 10$. A similar regression on the kurtosis of the distribution yields $\overline{\xi_4} \propto \overline{\xi_2}^{3.03 \pm 0.18}$ over essentially the same range in scales. Thus the data are consistent with the skewness and kurtosis being simply proportional to the square and the cube of the variance, respectively, both in the weakly and strongly non-linear regimes. Following our preliminary announcement of this result (Bouchet *et al.* 1992a), Gaztañaga (1992) analyzed the CfA and SSRS redshift surveys using similar techniques, and found very similar results. We looked directly at the ratio $S_3 = \overline{\xi_3}/\overline{\xi_2}^2$, and found that it varies only weakly, if at all, with scale; there is no theoretical reason to expect this to hold from weakly to strongly non-linear scales (although $N$-body simulations show similar behavior, Bouchet & Hernquist 1992; Suto 1993; Lucchin *et al.* 1993; D. Weinberg, private communication). The data are consistent with $S_3 = 1.5 \pm 0.5$. This value is smaller than that inferred from small scale measurements on optically selected samples, which probably reflects the weaker sampling of dense cluster cores by *IRAS* galaxies than optically selected galaxies. If linear biasing is assumed, the value of the biasing parameter is constrained between 1.6 and 3.2 (one sigma), for a power spectrum index $n = -1.4$, but the linear biasing model is probably not applicable when looking at statistics which measure the *asymmetry* of the density distribution. A similar analysis at the next order gives $S_4 = 4.4 \pm 3.7$; again, the dependence on scale is at most rather weak. Lahav *et al.* (1993) suggest that the constancy of $S_3$ with scale is partially due to redshift space distortions; Matsubara & Suto (1993) indeed find that $S_3$ grows with variance in analyses of $N$-body simulations evaluated in real space, but that in redshift space, it is much closer to constant.

We have tested the robustness of these results to the treatment of the clusters. *IRAS* galaxies are known to give systematically lower estimates of the density of cluster cores than do optically selected galaxies; "correction" for this effect greatly increases the observed strength of the correlations, and causes $S_3$ and $S_4$ to increase somewhat as a function of $\overline{\xi_2}$ in the transition between the weakly and the strongly non-linear regime, rather than staying constant. The boosting affects higher-order correlations more than those of lower-order, which means than the value of $S_3$ derived implies a lower value of the bias parameter than that found without biasing, $0.70 < b < 1.18$. However, this non-linear transformation of the galaxy density field introduces curvature terms into the relation between $S_3$ and bias. Moreover, it is not clear if *IRAS* galaxies or optically selected galaxies are a more faithful tracer of the underlying mass distribution. Finally, the boosting is done in a rather ad-hoc way, and is unlikely to reproduce all the features of counts in cells done from optically selected samples of galaxies; comparable analyses on large optically selected samples are needed to explore this further (cf., Gaztañaga 1992).

The astonishingly good agreement found here between the observed scalings, and those predicted by second-order perturbation theory given Gaussian initial conditions, appears to be a strong confirmation of the random-phase hypothesis. However, we have not explicitly tested non-Gaussian models; Luo & Schramm (1993) argue that non-linear evolution of initially non-Gaussian models make them appear remarkably like their Gaussian counterparts in the scalings of the various moments of the counts in cells. $N$-body simulations of non-Gaussian models, along the lines of Weinberg & Cole (1992) are needed to tell if the observed constancy of $S_3$ and $S_4$ can be used to rule out classes of such models (see Coles *et al.* 1993 for preliminary investigations along these lines). It is remarkable that the constancy of $S_3$ predicted from second-order perturbation theory holds true even in the non-linear regime ($\overline{\xi_2} \approx 10$). On the other hand, this is exactly what was found in $N$-body simulations of various cosmological scenarios by Bouchet *et al.* (1991) and Bouchet & Hernquist (1992), results that have been confirmed and extended by Lucchin *et al.* (1993) and Weinberg (private communication). However, an explanation based on dynamics continues to elude us. Moreover, Colombi *et al.* (1993a) find constancy of the $S_N$ in their $N$-body simulations only *after* finite volume corrections, if the simulation volume is too small; since we find the $S_N$ to be independent of scale without any such corrections, we may indeed have sampled volumes large enough not to be grossly affected at the scales we probed.

Szalay (1988) shows that if the biasing scheme is non-linear (that is, if there is a non-linear relation between the mass and galaxy density fields), then one generically expects a relation between the two and three-point correlation functions involving a cubic term, leading to a a dependence of $S_3$ on $\overline{\xi_2}$. However, this is a statement about initial conditions: non-linear evolution can quickly erase this cubic term (Gott, Bin, & Park 1991), and thus we cannot put constraints on the non-linearity of the biasing scheme. Moreover, Fry & Gaztañaga (1993) show that the scaling relations are preserved for *local* non-linear biasing schemes.

We found that the void probability function scales according to the prediction of the scale-invariant model, in which all $S_N$ are constant. This scaling is highly non-trivial, since we are probing the non-perturbative regime in which correlations of all orders are important. However, we only probe the dilute regime, for which the number of clustered particles above the mean is not much larger than unity. This limitation is due to the sparse sampling of the galaxy distribution by *IRAS* galaxies, and prevents study of further predictions of the scale invariant models, or discrimination between various competing models. Thus sparse sampling strategies (Kaiser 1986), while optimal for determining $\xi_2$ on large scales, are not appropriate for determining higher-order moments of the galaxy distribution. It also precludes using the overall shape of the counts to distinguish between the various existing theoretical models. Discrepancies are seen between the counts in cells and various such models in the densest subsample probed, but finite volume effects do not allow us to conclude firmly that these models do not fit the data.

What are the prospects for the future? It would be fascinating to extend the analysis discussed here to the highly non-linear regime, which will require densely sampled surveys of galaxies. Three surveys in progress come to mind. The Center for Astrophysics survey to $m_B = 15.5$ is nearing completion (cf., Geller & Huchra 1989), and certain aspects of the void distribution have already been discussed (Vogeley, Geller, & Huchra 1991). Davis *et al.* (1992) are finishing a magnitude-limited redshift survey of optically selected galaxies covering 65% of the sky, which will probe the densely sampled limit as well as the CfA survey (cf., Alimi *et al.* 1990), but in a much larger volume. Finally, Saunders *et al.* (1992) are in the process of obtaining redshifts for all *IRAS* galaxies to the flux limit of the *IRAS* Point Source Catalog, which should result in an appreciably denser sample than the one used in the present analysis. Thus we can look forward in a few years to testing if the scale-invariant forms, which seem to hold so well in $N$-body models, adequately describe the real universe.

**Acknowledgments:** We would like to thank S. Colombi, R. Juszkiewicz, R. Schaeffer, and D. Weinberg for many discussions on the general issue of counts in cells. MAS is supported at the IAS under NSF grant # PHY92-45317, and grants from the W.M. Keck Foundation and the Ambrose Monell Foundation. MD's research is supported in part by NASA grant NAG5-1360 and NSF grant AST-8915633. FRB would like to thank the hospitality of the Institute of Advanced Study in Princeton, where this work was completed.

# Appendix: Another Approach to $\overline{\xi_N}$

The correlation functions can be derived from the counts in cells data in a different way from that described in the main text. Let us define

$$Z_i \equiv \sum_{N=i}^{\infty} \overline{\xi_N} \frac{(-1)^i (-\overline{N})^{N-i}}{(N-i)!} \quad , \tag{31}$$

where, for convenience, we will set $\overline{\xi_0} \equiv 0$ and $\overline{\xi_1} \equiv 1$. In the limit $\overline{N} \ll 1$, we clearly have

$$Z_i \approx (-1)^i \overline{\xi_i} + \mathcal{O}(\overline{N}) \quad . \tag{32}$$

Also note that $\partial Z_N / \partial \overline{N} = Z_{N+1}$. Now we use the convenient fact that $P_0 = \exp Z_0$, and the derivative rule in Eq. 4 relating $P_N$ to $P_0$ to find:

$$P_1 = -\overline{N} Z_1 P_0 \quad , \tag{33a}$$

$$P_2 = \frac{\overline{N}^2}{2} (Z_2 + Z_1^2) P_0 \quad , \tag{33b}$$

$$P_3 = \frac{-\overline{N}^3}{6} (Z_3 + 3 Z_2 Z_1 + Z_1^3) P_0 \quad , \tag{33c}$$

$$P_4 = \frac{\overline{N}^4}{24} (Z_4 + 4 Z_1 Z_3 + 6 Z_1^2 Z_2 + 3 Z_2^2 + Z_1^4) P_0 \quad , \tag{33d}$$

and so on. Using the small $\overline{N}$ limit above, these can be turned directly into expressions for $\overline{\xi_N}$. Thus, measuring the $P$'s directly leads to estimates of the $\overline{\xi_N}$'s. We have found this process to be very unstable for large values of $N$, but using $P_2$ and $P_3$, we find reasonable measurements of $\overline{\xi_2}$ and $\overline{\xi_3}$. Fig. 18 shows this comparison: the solid squares connected by solid lines are the average correlation functions, taken from Fig. 3 and 5 above. The open circles connected by dashed lines are derived from $P_2$ and $P_3$ using the volume-limited sample to 7600 km s$^{-1}$ and Eq. 33$b$ and 33$c$, respectively. For this subsample, $\overline{N} < 0.1$ for the range of values of $\ell$ for which $\overline{\xi_N}$ is derived from this formalism, thus the expansion above is valid. The agreement is good, but because Eqs. 16 – 18 are exact, while Eq. 32 is an approximation, the former are more robust. Moreover, one gets positive results for the $\overline{\xi_N}$ over only a narrow range of scales, as Fig. 18 shows. Thus one can determine values of $\overline{\xi_4}$ from $P_4$ at only two scales! Thus this method is inferior to the determination of the $\overline{\xi_N}$ from moments.

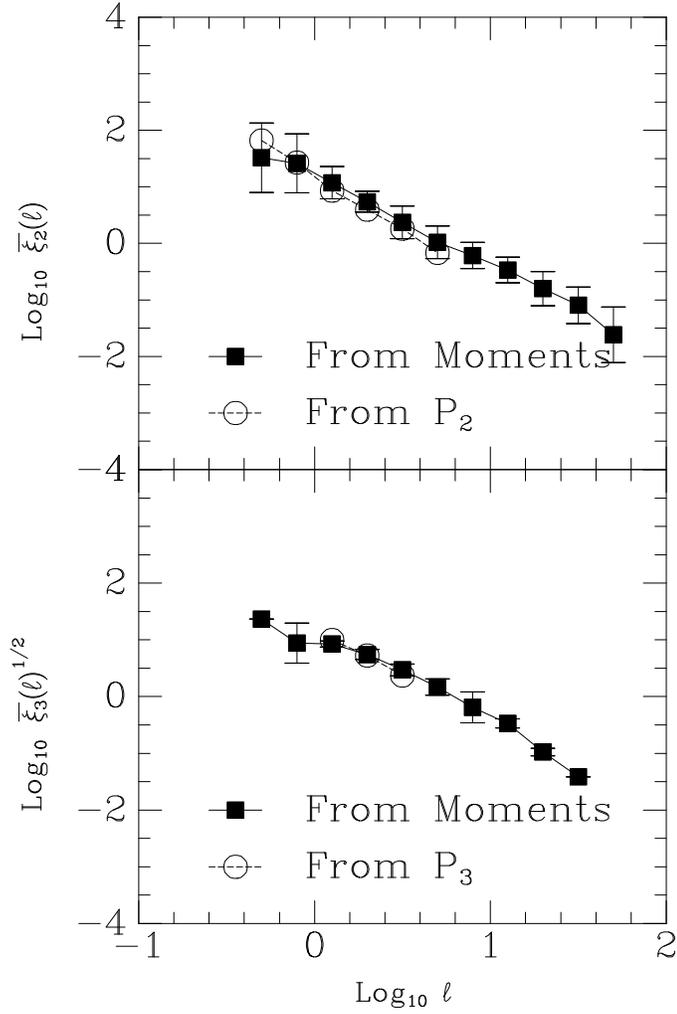

Figure 18: The two- and three-body correlation functions $\overline{\xi_2}(\ell)$ (top), and $\overline{\xi_3}(\ell)$ (bottom) as determined from Eq. 32 (open symbols). They agree very well with the determinations from the moments of the counts in cells (filled symbols), but only a much narrower range of scales can be analyzed in such a way.

Table 1

Volume-limited Samples Used

| Maximum Radius[a] | $H_0 r$[a,b] | # Galaxies | Log$_{10}$ Luminosity[c] | # Random Galaxies[d] |
|---|---|---|---|---|
| 2400 | 2386 | 444 | 8.95 | 16 |
| 3900 | 3862 | 609 | 9.36 | 23 |
| 4800 | 4743 | 748 | 9.54 | 28 |
| 6000 | 5911 | 877 | 9.73 | 32 |
| 7600 | 7458 | 788 | 9.92 | 29 |
| 9500 | 9280 | 699 | 10.12 | 26 |
| 12,000 | 11651 | 572 | 10.32 | 21 |
| 15,000 | 14460 | 428 | 10.53 | 16 |
| 19,000 | 18142 | 302 | 10.70 | 11 |
| 24,000 | 22648 | 203 | 10.89 | 8 |

[a] in km s$^{-1}$

[b] calculated from Equation 13 with $\Omega_0 = 1$.

[c] Minimum luminosity of volume-limited subsample. $L = 4\pi r^2 \nu f_\nu$ at 60$\mu$m, with $H_0 = 100$ km s$^{-1}$ Mpc$^{-3}$. Units are solar luminosities.

[d] $|b| > 5°$ only

Table 2

Least-Square Fits to Correlation Functions $\overline{\xi_N}$ for $\Omega = 1.0$; Clusters Not Collapsed

| Maximum Radius | $A_N$[a] | $B_N$ | $C_N$[b] | $D_N$ | $S_N$[c] |
|---|---|---|---|---|---|
| | | | $N = 2$ | | |
| 2400 | $0.86 \pm 0.04$ | $-1.49 \pm 0.08$ | — | — | — |
| 3900 | $1.04 \pm 0.03$ | $-1.53 \pm 0.05$ | — | — | — |
| 4800 | $1.06 \pm 0.06$ | $-1.57 \pm 0.08$ | — | — | — |
| 6000 | $1.18 \pm 0.06$ | $-1.60 \pm 0.08$ | — | — | — |
| 7600 | $1.16 \pm 0.07$ | $-1.68 \pm 0.09$ | — | — | — |
| 9500 | $1.44 \pm 0.03$ | $-1.78 \pm 0.04$ | — | — | — |
| 12000 | $1.42 \pm 0.02$ | $-1.74 \pm 0.03$ | — | — | — |
| 15000 | $1.79 \pm 0.14$ | $-2.02 \pm 0.15$ | — | — | — |
| 19000 | $1.39 \pm 0.34$ | $-0.73 \pm 0.29$ | — | — | — |
| 24000 | $1.39 \pm 0.20$ | $-1.42 \pm 0.17$ | — | — | — |
| All | $1.17 \pm 0.05$ | $-1.59 \pm 0.06$ | — | — | — |
| | | | $N = 3$ | | |
| 2400 | $1.00 \pm 0.07$ | $-1.72 \pm 0.15$ | $0.01 \pm 0.06$ | $2.33 \pm 0.07$ | $1.52 \pm 0.71$ |
| 3900 | $1.20 \pm 0.06$ | $-1.55 \pm 0.09$ | $0.03 \pm 0.02$ | $1.92 \pm 0.04$ | $1.93 \pm 0.28$ |
| 4800 | $1.14 \pm 0.08$ | $-1.44 \pm 0.12$ | $0.21 \pm 0.05$ | $1.86 \pm 0.09$ | $1.59 \pm 0.37$ |
| 6000 | $1.29 \pm 0.13$ | $-1.59 \pm 0.16$ | $0.23 \pm 0.05$ | $1.98 \pm 0.08$ | $1.85 \pm 0.43$ |
| 7600 | $1.25 \pm 0.08$ | $-1.68 \pm 0.09$ | $0.12 \pm 0.06$ | $1.86 \pm 0.07$ | $1.26 \pm 0.29$ |
| 9500 | $1.32 \pm 0.11$ | $-1.58 \pm 0.15$ | $0.08 \pm 0.06$ | $2.01 \pm 0.13$ | $1.25 \pm 0.32$ |
| 12000 | $1.56 \pm 0.18$ | $-1.85 \pm 0.18$ | $0.07 \pm 0.11$ | $2.07 \pm 0.22$ | $1.17 \pm 0.41$ |
| All | $1.11 \pm 0.05$ | $-1.46 \pm 0.06$ | $0.15 \pm 0.05$ | $1.96 \pm 0.06$ | $1.53 \pm 0.48$ |
| | | | $N = 4$ | | |
| 2400 | $1.13 \pm 0.10$ | $-2.02 \pm 0.24$ | $-0.41 \pm 0.02$ | $4.40 \pm 0.38$ | $2.92 \pm 2.36$ |
| 3900 | $1.23 \pm 0.16$ | $-1.55 \pm 0.23$ | $-0.50 \pm 0.17$ | $2.94 \pm 0.31$ | $3.99 \pm 2.61$ |
| 4800 | $1.49 \pm 0.15$ | $-1.88 \pm 0.20$ | $0.38 \pm 0.11$ | $3.48 \pm 0.24$ | $2.99 \pm 1.49$ |
| 6000 | $1.19 \pm 0.20$ | $-1.25 \pm 0.26$ | $0.67 \pm 0.18$ | $2.52 \pm 0.41$ | $5.24 \pm 2.65$ |
| 7600 | $1.16 \pm 0.12$ | $-1.39 \pm 0.15$ | $0.33 \pm 0.22$ | $2.83 \pm 0.45$ | $2.51 \pm 1.66$ |
| 12000 | $2.0 \pm 0.08$ | $-1.97 \pm 0.08$ | $1.13 \pm 0.03$ | $3.24 \pm 0.10$ | $12.7 \pm 2.87$ |
| All | $1.15 \pm 0.09$ | $-1.35 \pm 0.12$ | $0.46 \pm 0.09$ | $3.03 \pm 0.18$ | $4.39 \pm 3.68$ |
| | | | $N = 5$ | | |
| 2400 | $0.92 \pm 0.13$ | $-1.54 \pm 0.68$ | $-0.45 \pm 0.16$ | $4.8 \pm 2.40$ | $2.43 \pm 2.95$ |
| 3900 | $1.58 \pm 0.03$ | $-1.88 \pm 0.06$ | $0.88 \pm 0.06$ | $5.0 \pm 0.16$ | $20.4 \pm 14.9$ |
| 4800 | $1.30 \pm 0.01$ | $-1.39 \pm 0.03$ | $1.40 \pm 0.04$ | $3.7 \pm 80.10$ | $9.24 \pm 1.54$ |
| 6000 | $1.58 \pm 0.07$ | $-1.53 \pm 0.09$ | $1.39 \pm 0.06$ | $4.3 \pm 70.22$ | $28.4 \pm 7.14$ |
| All | $1.05 \pm 0.10$ | $-0.92 \pm 0.16$ | $1.27 \pm 0.18$ | $3.28 \pm 0.38$ | $22.1 \pm 25.3$ |

[a] Fit to $1/(N-1) \log_{10} \overline{\xi_N}(\ell) = A_N + B_N \log_{10} \ell$.
[b] Fit to $\log_{10} \overline{\xi_N}(\ell) = C_N + D_N \log_{10} \overline{\xi_2}(\ell)$.
[c] Average of $\overline{\xi_N}/\overline{\xi_2}^{N-1}$ in the range $0.1 < \overline{\xi_2} < 10$.

# References

Alimi, J.-M., Blanchard, A., & Schaeffer, R., 1990, ApJ, 349, L5
Balian, R., & Schaeffer, R., 1988, ApJ, 335, L43
Balian, R., & Schaeffer, R., 1989a, A&A, 220, 1
Balian, R., & Schaeffer, R., 1989b, A&A, 226, 373
Bernardeau, F., 1992, ApJ, 392, 1
Bouchet, F.R., Davis, M., & Strauss, M.A., 1992a, The Distribution of Matter in the Universe, ed. G.A. Mamon & D. Gerbal (Meudon: Observatoire de Paris), p. 287
Bouchet, F.R., & Hernquist, L., 1992, ApJ, 400, 25
Bouchet, F.R., Juszkiewicz, R., Colombi, S., & Pellat, R., 1992b, ApJ, 394, L5
Bouchet, F.R., Schaeffer, R., & Davis, M., 1991, ApJ, 383, 19 (BSD)
Carruthers, P., & Shih, C.C., 1983, Phys. Lett. B, 127, 242
Coles, P., & Frenk, C.S., 1991, MNRAS, 253, 727
Coles, P., & Jones, B., 1991, MNRAS, 248, 1
Coles, P., Moscardini, L., Lucchin, F., Matarrese, S., & Messina, A., 1993, preprint
Colombi, S., & Bouchet, F.R., 1992, The Distribution of Matter in the Universe, ed. G.A. Mamon & D. Gerbal (Meudon: Observatoire de Paris), p. 273
Colombi, S., Bouchet, F.R., & Schaeffer, R., 1993a, A&A, in press
Colombi, S., Bouchet, F.R., & Schaeffer, R., 1993b, in preparation
Davis, M., Geller, M.J., & Huchra, J.P., 1978, ApJ, 221, 1
Davis, M., Meiksin, A., Strauss, M.A., da Costa, N., & Yahil, A., 1988, ApJ, 333, L9
Davis, M., & Peebles, P.J.E., 1977, ApJS, 34, 425
Davis, M., Strauss, M.A., & Yahil, A., 1991, ApJ, 372, 394
Davis, M. *et al.*, 1992, in preparation
de Lapparent, V., Geller, M.J., & Huchra, J.P., 1986, ApJ, 302, L1
Efstathiou, G., Kaiser, N., Saunders, W., Lawrence, A., Rowan-Robinson, M., Ellis, R.S., & Frenk, C.S., 1990, MNRAS, 247, 10p
Fisher, K.B., 1992, PhD. Thesis, University of California, Berkeley
Fisher, K.B., Davis, M., Strauss, M.A., Yahil, A., & Huchra, J.P., 1993a, ApJ, 402, 44
Fisher, K.B., Davis, M., Strauss, M.A., Yahil, A., & Huchra, J.P., 1993b, submitted to MNRAS
Fisher, K.B., Strauss, M.A., Davis, D., Yahil, A., & Huchra, J.P., 1992, ApJ, 389, 188
Fry, J., 1984, ApJ, 279, 499
Fry, J., 1986, ApJ, 306, 358
Fry, J.N, & Gaztañaga, E., 1993, preprint
Fry, J.N., Melott, A.L., & Shandarin, S.F., 1993, preprint
Fry, J., & Peebles, P.J.E., 1978, ApJ, 221, 19
Gaztañaga, E., 1992, ApJ, 398, L17
Gaztañaga, E., & Yokohama, J., 1993, ApJ, 403, 450
Geller, M.J., & Huchra, J.P., 1989, Science, 246, 897
Goroff, M.H., Grinstein, B., Rey, S.-J., & Wise, M.B., 1986, ApJ, 311, 6
Gott, J.R., Bin, G., & Park, C., 1991, ApJ, 383, 90
Groth, E., & Peebles, P.J.E., 1977, ApJ, 217, 385
Juszkiewicz, R., & Bouchet, F.R., 1992, The Distribution of Matter in the Universe, ed. G.A. Mamon & D. Gerbal (Meudon: Observatoire de Paris), p. 301
Juszkiewicz, R., Bouchet, F.R., & Colombi, S., 1993, ApJ, in press
Juszkiewicz, R., Weinberg, D., Amsterdamski, P., Chodorowski, M., & Bouchet, F., 1993, in preparation
Kaiser, N., 1986, MNRAS, 219, 785

Kirshner, R.P., Oemler, A., Schechter, P.L., & Shectman, S.A., 1987, ApJ, 314, 493

Lahav, O., Itoh, M., Inagaki, S., & Suto, Y., 1993, ApJ, 402, 387

Little, B., Weinberg, D.H., & Park, C., 1991, MNRAS, 253, 295

Lucchin, F., Matarrese, S., Melott, A. L., & Moscardini, L., 1993, preprint

Luo, X., & Schramm, D., 1993, ApJ, in press

Maddox, S.J., Efstathiou, G., Sutherland, W., & Loveday, J., 1990, MNRAS, 242, 43P

Matsubara, T., & Suto, Y., 1993, preprint

Maurogordato, S., Schaeffer, R., & da Costa, L.N., 1992, ApJ, 390, 17

Meiksin, A., Szapudi, I., & Szalay, A., 1992, ApJ, 394, 87

Peebles, P.J.E., 1980, The Large Scale Structure of the Universe (Princeton: Princeton University Press)

Peebles, P.J.E., 1993, Principles of Physical Cosmology (Princeton: Princeton University Press)

Peebles, P.J.E., & Groth, E.J., 1976, A&A, 53, 131

Pietronero, P., 1987, Physica, 144A, 257

Rowan-Robinson, M. *et al.*, 1990, MNRAS, 247, 1 (QDOT)

Saslaw, W.C., 1985, ApJ, 297, 49

Saslaw, W.C., 1989, ApJ, 341, 588

Saslaw, W.C., & Hamilton, A.J.S., 1984, ApJ, 276, 13

Saunders, W., Frenk, C., Rowan-Robinson, M., Efstathiou, G., Lawrence, A., Kaiser, N., Ellis, R., Crawford, J., Xia, X.-Y., & Parry, I., 1991, Nature, 349, 32

Saunders, W. *et al.*, 1992, in preparation

Schaeffer, R., 1984, A&A, 134, L15

Sharp, N., Bonometto, S.A., & Lucchin, F., 1984, A&A, 130, 79

Strauss, M.A., Davis, M., Yahil, A., & Huchra, J.P., 1990, ApJ, 361, 49

Strauss, M.A., Davis, M., Yahil, A., & Huchra, J.P., 1992a, ApJ, 385, 421

Strauss, M.A., Huchra, J.P. Davis, M., Yahil, A., Fisher, K.B., & Tonry, J.T., 1992b, ApJS, 83, 29

Suto, Y., 1993, ApJ, 404, L1

Suto, Y., Itoh, M., & Inagaki, S., 1990, ApJ, 350, 492

Szalay, A.S., 1988, ApJ, 333, 21

Szapudi, I., & Szalay, A.S., 1993, preprint

Szapudi, I., Szalay, A.S., & Boshan, P., 1992, ApJ, 390, 350

Vogeley, M.S., Geller, M.J., & Huchra, J.P., 1991, ApJ, 382, 44

Weinberg, D., 1989, PhD. Thesis, Princeton University

Weinberg, D., & Cole, S., 1992, MNRAS, 259, 652

White, S.D.M., 1979, MNRAS, 186, 145

Yahil, A., Strauss, M.A., Davis, M., & Huchra, J.P., 1991, ApJ, 372, 380

Yahil, A., Tammann, G., & Sandage, A., 1977, ApJ, 217, 903